%% file: main.tex
\documentclass[a4paper,11pt]{article}
\pdfoutput=1

\usepackage{jcappub}

\usepackage[T1]{fontenc} 

\usepackage{ulem} 

\title{\boldmath Characterization of the polarized synchrotron emission from Planck and WMAP data}

\author{F. A. Martire,}
\author{R. B. Barreiro}
\author{and E. Mart\'{\i}nez-Gonz\'alez}

\affiliation{Instituto de F\'isica de Cantabria, CSIC-Universidad de Cantabria,\\ 
Avda. de los Castros s/n, E-39005 Santander, Spain}

\emailAdd{martire@ifca.unican.es}
\emailAdd{barreiro@ifca.unican.es}
\emailAdd{martinez@ifca.unican.es}

\abstract{
The purpose of this work is to characterize the diffuse Galactic polarized synchrotron, which is the dominant CMB foreground emission at low frequency. We present EE, BB, and EB power spectra estimated from polarization frequency maps at 23 and 30 GHz as observed respectively by the \textit{WMAP} K-band and the \textit{Planck} lowest frequency channel, for a set of six sky regions covering from 30\% to 94\% of the sky. We study the synchrotron polarization angular distribution and spectral energy distribution (SED) by means of the so-called pseudo-$C_\ell$ formalism, provided by the \texttt{NaMaster} package, in the multipole interval 30 $\leq$ $\ell$ $\leq$ 300. Best results are obtained cross-correlating \textit{Planck} and \textit{WMAP} data. The EE and BB angular power spectra show a steep decay of the spectral amplitude as a function of multipole, approximated by a power law $C^{EE,BB} \propto \ell^{\alpha_{EE,BB}}$, with $\alpha_{EE}= -2.95\pm0.04$ and $\alpha_{BB}=-2.85\pm0.14$. The B/E power asymmetry is proved with a B-to-E ratio, computed as the amplitude ratio at the pivot multipole $\ell = 80$, of 0.22$\pm$0.02. The EB cross-component is compatible with zero at 1$\sigma$, with an upper constraint on the EB/EE ratio of 1.2\% at the 2$\sigma$ level. We show that the EE and BB power-law model with null EB cross-correlation describes reasonably well the diffuse synchrotron polarization emission for the full sky if the bright Galactic center and point sources are masked.
The recovered SED shows power-law spectral indices $\beta_{EE}= -3.00 \pm 0.10$ and $\beta_{BB} = -3.05\pm0.36$ compatible between themselves, in the frequency range 23–30 GHz. Results also seem to indicate that the SED gets steeper from low to high Galactic latitude.
}

\begin{document}
\maketitle
\flushbottom

\input{sections/section01.tex}
\input{sections/section02.tex}
\input{sections/section03.tex}

\input{sections/section04.tex}
\input{sections/section05.tex}
\input{sections/section06.tex}

\acknowledgments
The authors would like to thank the Spanish Agencia Estatal de Investigaci\'on (AEI, MICIU) for the financial support provided under the projects with references PID2019-110610RB-C21, ESP2017-83921-C2-1-R and AYA2017-90675-REDC, co-funded with EU FEDER funds, and acknowledge  support from Universidad de Cantabria and Consejer{\'\i}a de Universidades,  Igualdad,  Cultura  y  Deporte  del  Gobierno de Cantabria via the “Instrumentaci{\'o}n y ciencia de datos para sondear la naturaleza del universo” project as well as from the  Unidad de Excelencia Mar{\'\i}a de Maeztu (MDM-2017-0765). FAM is supported by a fellowship funded by the Unidad de Excelencia María de Maeztu. The authors also thank J.D. Bilbao-Ahedo and M. López-Caniego for useful discussions. We acknowledge Santander Supercomputacion support group at the University of Cantabria who provided access to the supercomputer Altamira Supercomputer at the Institute of Physics of Cantabria (IFCA-CSIC), member of the Spanish Supercomputing Network, for performing simulations/analyses. This research used resources of the National Energy Research Scientific Computing Center (NERSC), a U.S. Department of Energy Office of Science User Facility located at Lawrence Berkeley National Laboratory, operated under Contract No. DE-AC02-05CH11231.
Some of the presented results are based on observations obtained with \textit{Planck}\footnote{http://www.esa.int/Planck}, an ESA science mission with instruments and contributions directly funded by ESA Member States, NASA, and Canada.
We also acknowledge the Legacy Archive for Microwave Background Data Analysis (LAMBDA). Support for LAMBDA is provided by the NASA oﬃce of Space Science.
Some of the results in this work have been derived using the \texttt{HEALPix} \cite{healpy1}, \texttt{NaMaster} \cite{namaster}, \texttt{PySM} \cite{PySM}, \texttt{Matplotlib} \cite{pltpy}, \texttt{NumPy} \cite{NumPy} and \texttt{SciPy} \cite{SciPy} Python packages. 
We also thank an anonymous referee for the useful comments.

\appendix
\input{sections/appendix01.tex}
\input{sections/appendix02.tex}

\input{sections/appendix03.tex}

\newpage

\bibliographystyle{JHEP.bst}
\bibliography{main.bib}

\end{document}

%% file: sections/section01.tex
\section{Introduction} 
\label{sec:01}
In modern Cosmology, the cosmic microwave background (CMB) plays a fundamental role describing the physics of our Universe, its structure and how it evolved in time. 
Among many other remarkable experiments, the CMB anisotropies have been observed with the Wilkinson Microwave Anisotropy Probe (\textit{WMAP}) \cite{Bennett_2013} and, with increasing precision, by the \textit{Planck} satellite \cite{planckLegacy}.
Further cosmological progress will be made by precisely measuring the polarization CMB anisotropies. It has been proved that primordial gravitational waves (GWs) 
would leave a unique imprint in the so-called B-mode of CMB polarization \cite{HuWhite_1997}. Therefore, its detection would provide a strong evidence in support of some theories on the very early stage of the Universe \cite{HuWhite_1996,HuWhite_1996_2,SpergelZald_1997}. \\
However, mixed with the cosmological signal, CMB observations also contain different astrophysical emissions, usually called \emph{CMB foregrounds}.
The accuracy of the CMB measurements thus depends critically on the foregrounds removal process, which is commonly done by a component separation algorithm; for the power spectrum and cosmological parameters estimation, marginalization over some foreground parametrization is usually done \cite{CompSep2016,akrami2020planck}. These methods require different levels of prior knowledge of the foregrounds, in both intensity and polarization, on their spatial fluctuations, and most importantly, spectral information \cite{Leach_2008}.

The foregrounds are usually characterized as Galactic or extragalactic, and compact or diffuse, which can dominate at the lower or higher frequencies of interest.
In polarization, there are currently two known important diffuse Galactic foregrounds: synchrotron at low frequency and thermal dust emission at high frequency, which dominate respectively roughly below and above 100~GHz. The synchrotron radiation is generated by relativistic cosmic ray electrons accelerating around the Galactic magnetic field, which spiral around the field lines emitting radiation. The thermal dust emission comes, instead, from interstellar dust grains aligned with the Galactic magnetic field. Besides these two main components, it is worth mentioning other two important foreground sources: free-free radiation, which is emitted by free electrons interacting with ionised gas, and, Anomalous Microwave Emission (AME), explained as rotational emission from ultra-small dust grains (spinning dust) \cite{Dickinson_2018}. However, these two foregrounds are not considered in this work because free-free is intrinsically unpolarized and AME is expected to be very weakly polarized, with the best current constraints imposed for the molecular cloud complex W43 with \textit{QUIJOTE} ($\Pi_{AME}$ <0.39 at 95 per cent CL at 17 GHz) and \textit{WMAP} ($\Pi_{AME}$<0.22 at 95 per cent CL at 41 GHz) data \cite{G_nova_Santos_2016}. \\
The characterization of the thermal dust polarization, at intermediate and high Galactic latitudes, has been carried out with the \textit{Planck} data at high frequencies \cite{planckDust2016}. Measurements were sensitive enough to determine that: the thermal dust polarization frequency dependence is consistent with a modified black-body, the power spectra are well described by the power law $C_\ell^{EE,BB}\propto\ell^{-2.42\pm 0.02}$, and the B-to-E ratio is $C_\ell^{BB}/C_\ell^{EE} \approx 0.5$ in the multipole range 40 $\leq$ $\ell$ $\leq$ 600. Moreover, the dust EB signal is consistent with zero and the EB/EE ratio is smaller than 3\% \cite{planckDust}.

The sensitivity at low frequency of current space-based observations, \textit{WMAP} and \textit{Planck}, does not allow a good characterization of the synchrotron polarization signal at intermediate and high Galactic latitudes.
Thus, observations from ground-based experiments at lower frequencies are very useful to trace the synchrotron emission due to their high signal-to-noise ratio, such as the S-band Polarization All Sky Survey (\textit{S-PASS}) \cite{Carretti_2019} at 2.3 GHz, the C-band All Sky Survey (\textit{C-BASS}) \cite{C-BASS_2019} at 5 GHz or the \textit{QUIJOTE-MFI} Northern sky survey \cite{quijote12}, covering frequencies from 11 to 19 GHz and whose first release is expected in the next months. Note that although lower frequencies present higher signal to noise ratios, they can also be significantly affected by Faraday rotation effects, which can depolarize the synchrotron emission, especially around the Galactic plane \cite{FuskelandSynch}. Conversely, for higher frequencies (such as those from \textit{QUIJOTE-TGI}), the synchrotron amplitude is lower but they are in the safer side with respect to depolarization, while also being closer to the frequencies of interest for CMB observations.
In this work, we characterize the diffuse synchrotron polarization analyzing the observations of \textit{WMAP} K-band and \textit{Planck} 30~GHz frequency channels focusing on the intermediate and low Galactic latitudes. This is done by considering the sky regions allowed by a set of customized masks that remove the Galactic center as well as compact sources, but keeping other bright regions of synchrotron emission, trying to maximise the diffuse synchrotron signal-to-noise ratio. Results are given for both \textit{Planck} and \textit{WMAP} data independently, as well as cross-correlating the data sets from both experiments, the latter constituting our reference results.

%% file: sections/section02.tex
\section{Data and Simulations}
\label{sec:02}
\subsection{Data}
For our analysis, we will use \textit{Planck} and \textit{WMAP} data.
\textit{Planck} was a space-based experiment consisting of two instruments, the Low Frequency Instrument (LFI) and the High Frequency Instrument (HFI), observing both the total intensity and polarization of sky photons, and covering a wide frequency range from 30 to 857 GHz with 9 frequency channels. In this work, we use the 2018 data release (PR3) \cite{planckLegacy}, obtained from the full set of observations, focusing on the lowest channel at central frequency 28.4~GHz. In Appendix \ref{appendix:02,subsection:01}, we use instead the 2020 \textit{Planck} release (PR4), computed with the \texttt{NPIPE} processing pipeline \cite{Npipe}, in order to test the consistency with our main results. 
\textit{Planck} data were downloaded from the \textit{Planck} Legacy Archive\footnote{pla.esac.esa.int} (PLA) and then downgraded down to the pixel resolution corresponding to the $N_{side} = 512$ \texttt{Healpix} parameter. Note that the resolution of the PR3 (PR4) 30 GHz \textit{Planck} map corresponds to an effective beam of FWHM = 32.39 (31.5) arcminutes. \\
\textit{WMAP} was also a space-based experiment, which observed the total intensity and polarization of the sky, using a narrower frequency range, from 23 to 94 GHz, with five frequency bands. In our analysis, we include the lowest frequency channel of the \textit{WMAP} dataset, namely the \textit{K}-band centered at 23 GHz \cite{Bennett_2013}, obtained from the 9-yr data release. All \textit{WMAP} products have been downloaded from LAMBDA\footnote{lambda.gsfc.nasa.gov/product/map} and have been analysed at their original resolution ($N_{side}=512$, with an effective beam of FWHM=0.88$^\circ$).

\begin{figure}[htbp]
\centerline{\includegraphics[scale=.25]{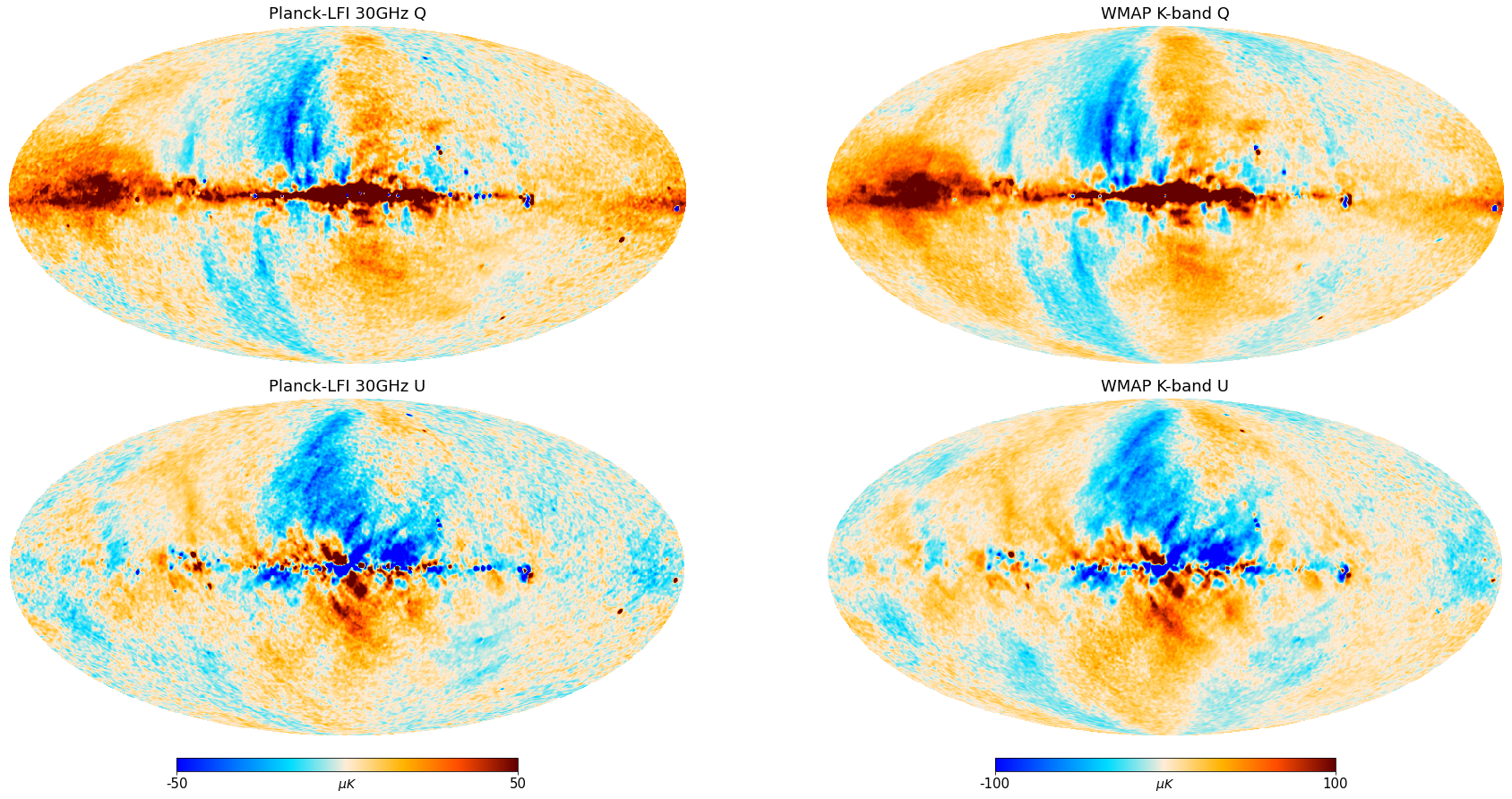}}
\caption{Left: Q (top) and U (bottom) polarization all-sky maps of the \textit{Planck} 30~GHz frequency channel \cite{planckLegacy}. Right: Same maps for the \textit{WMAP} K-band \cite{Bennett_2013}. All maps have a resolution of $1^\circ$.}
\label{fig:maps}
\end{figure}

In order to estimate the power spectra from only \textit{Planck} data, we cross-correlate the two half-ring 30 GHz maps, that are generated using only the first and the second halves of each pointing period, respectively. Using the cross-correlation of splits, rather than the auto-spectra from the full mission data, has the advantage of cancelling instrumental noise and reducing the effect of systematics. For the only-\textit{WMAP} analysis, we follow an analogous procedure and use as splits the co-added maps from 1 to 4 years on one side, and from 5 to 9 years on the other.
Power spectra results are also obtained from the cross-correlation of \textit{WMAP} and \textit{Planck} maps, using in this case the full-mission \textit{Planck} and the co-added nine-year \textit{WMAP} maps. By cross-correlating data from independent experiments, we can use directly the full data set rather than the splits, since the instrumental noise is uncorrelated and the effect of the systematics is also reduced. This also allows us to use the larger number of simulations which are available at the PLA for the \textit{Planck} full mission case with respect to the half-ring splits.
The full-mission \textit{Planck} maps have significantly lower noise than the nine-year \textit{WMAP} maps, however, the synchrotron brightness in the \textit{Planck} lowest frequency, at 28~GHz, is around half that in the \textit{WMAP} K-band, at 23 GHz, what ends up in very similar foreground signal-to-noise for both experiments. According to \cite{Planck2016Diff}, at a scale of $1^\circ$, the median (mean) signal-to-noise for \textit{WMAP} K-band is 2.47 (3.77) while for \textit{Planck} 30~GHz we have 2.64 (3.72). Nevertheless, each map is better in some sky regions because of the different scan strategies. Maps are shown in Figure \ref{fig:maps}.

\subsection{Simulations}

In order to estimate the errors of our set of power spectra and evaluate the goodness of fit, we use simulations. The simulations for the two instruments, both splits and full-mission, contain the sum of CMB, foregrounds and noise. A foreground simulation is generated with the Python Sky Model package\footnote{pysm3.readthedocs.io} (\texttt{PySM}) \cite{PySM}, currently used for the sky modeling of many CMB analysis, at \textit{Planck} and \textit{WMAP} frequencies with the considered resolutions\footnote{Note that this applies to the analyses including WMAP and/or PR3.
PR4, however, provides full simulations including foregrounds, that have been generated by evaluating the Commander sky model at the target frequency \cite{Npipe}.}. The chosen model includes two polarized components: thermal dust, modelled as a single-component modified black body (d1), and synchrotron emission, described as a power law scaling with a spatially varying spectral index (s1). AME and free-free emission are not included, since they are assumed to be unpolarised.
We recall that the PySM synchrotron model was obtained combining the first \textit{WMAP} polarization data with the Haslam total intensity map at 408 MHz \cite{Haslam1981}, including a model for the Galactic magnetic field \cite{Miville-Deschenes}.

\noindent Note that we are always using the same foreground model for all simulations. This procedure has also been followed in previous analyses \cite{planckDust, Minami_2020} due to the fact that foregrounds are deterministic and also to the difficulty of producing different realistic foreground models. However, this procedure does not include the uncertainties associated to the knowledge of the foreground model, which are difficult to quantify, what could lead to a covariance matrix somewhat underestimated. In order to test the effect of the foreground model on our results, we repeated the analysis for our reference case using an alternative PySM model ("d2s2") finding fully compatible results.\footnote{Moreover, the analysis carried out using PR4 also considers a different (although again deterministic) foreground simulation, providing again consistent results (see Appendix~\ref{appendix:02,subsection:01}). This further confirms the stability of our conclusions versus the considered foreground model.}
We remark that it is very important to include an estimation of the foreground emission, even if fixed for all simulations, since this introduces significant variance in the power spectra due to the presence of chance correlations between the foregrounds and the other components.

For the \textit{Planck} analysis, we use the CMB and noise PR3-2018 simulations (FFP10) at the 30~GHz channel, generated using the end-to-end simulation pipeline \cite{Planck_3}, provided by the \textit{Planck} Legacy Archive, which are also degraded to $N_{side}=512$, the resolution considered in our analysis. In particular, we use 600 FFP10 lensed CMB maps \cite{Planck2015_sim}, 300 full-mission LFI E2E simulations and 100 half-ring LFI E2E simulations per split, being the number of the noise simulations limited by the availability in the PLA. The LFI E2E simulations include noise and systematics due to realistic instrumental effects, which are then processed with the same algorithms as for the flight data.
For a consistency test, we also use the new full-mission and A/B-splits \textit{Planck} PR4 simulations, but further details are presented in Appendix \ref{appendix:02,subsection:01}. 

For the \textit{WMAP} analysis, we generate 300 CMB Gaussian realizations using the power spectra from the \textit{Planck} best-fit $\Lambda$CDM model \cite{planckLegacy} at the \textit{WMAP} K-band channel resolution, using the \texttt{Healpy} package\footnote{healpy.readthedocs.io} \cite{healpy1,healpy2}.
In addition, we generate a set of 600 noise simulations consistent with the full \textit{WMAP} 9-yr and 2 sets of 300 noise simulations consistent with the split data sets. The full-data noise simulations are obtained from the full pixel by pixel covariance matrix, while those for the splits are generated combining the single-year covariance matrices from year 1 to 4 on one side and from year 5 to 9 on the other. The single-year and the full-mission covariance matrices are provided as LAMBDA Products.

%% file: sections/section03.tex
\section{Masks}
\label{sec:03}

Although Galactic foregrounds studies usually focus on regions at high Galactic latitudes (since these are of greater interest for CMB analyses), these regions are very affected by noise in the \textit{WMAP} 23~GHz and \textit{Planck} 30~GHz polarised maps. Therefore, in order to have a higher signal-to-noise, our analysis will instead concentrate on low and intermediate latitudes, by constructing a set of customised masks with different sky fractions. For comparison, and in order to test the validity of our results also at high Galactic regions, the (almost) full-sky case will also be considered.
In particular, our masks are constructed as a combination of a Galactic mask (that removes the brightest Galactic centre), a point source mask and a series of polarization masks constructed by thresholding the total polarised intensity ($P = \sqrt{Q^2 + U^2}$) of the \textit{Planck} 30~GHz map. In this way, we end up with a set of five custom masks at intermediate and low Galactic latitude, which provide a useful sky fraction from 30 to 70\%. For completeness, we also consider a 94\% mask constructed simply combining the Galactic and the point sources masks.

\subsection{Galactic Mask}
The emission of the central part of the Galactic centre has a very complex behaviour and, therefore, can not be characterised with a simple power law model, as the one considered in this work. Therefore, it needs to be excluded from our analysis. For this we construct a customised Galactic mask in the following way. \\
First, as baseline, we exclude those pixels given by the 2015 Galactic plane mask (provided in the PLA) that leaves 97 per cent of the sky unmasked and that has been derived from \textit{Planck} higher frequency channels. Second, in order to adapt better our customised mask to the considered data maps, we also exclude those pixels with $P > 70~ \mu K$ in the \textit{Planck} 30 GHz channel and those with $P > 280~ \mu K$ in the \textit{WMAP} 23 GHz map, both smoothed at $5^{\circ}$. 
The thresholds were set independently for each map in order to select the best region around the Galactic centre, such that the results of the analyses were robust while discarding only a small fraction of the sky. \\
In order to regularise the boundaries, the resulting mask is then smoothed with a Gaussian beam with FWHM=$2^{\circ}$. All pixels with values $\leq 0.8$ are considered for the mask and, therefore, discarded from the analysis.
The final Galactic mask retains a sky fraction of $f_{sky}= 0.95$ (see grey Galactic region of Fig.~\ref{fig:masks}).

\subsection{Point Source Mask}
The previous Galactic mask is not enough to ensure the exclusion of all regions with complex emission that do not follow a simple model for the power spectra. Indeed, we have seen that very bright point sources, both Galactic and extragalactic, can have a significant effect at the spectra at all scales.
Therefore, we generate a point sources mask as a combination of \textit{Planck} and \textit{WMAP} polarization point source masks. \\
For \textit{Planck}, we use the mask for the 30 GHz polarised map of the SEVEM pipeline (one of the four component separation methods used by the \textit{Planck} Collaboration), which consists of 195 point sources that have polarization detection significance levels of 99\% or more \cite{akrami2020planck,FilteredFusion,mhw2}. The \textit{WMAP} point sources mask is generated from a point source catalog \cite{WMAP_PS} of 22 objects, where each source is detected in polarization with a significance level greater than 99.99\% in at least one \textit{WMAP} channel.
After combining these two masks, we smooth it with a Gaussian beam with FWHM=30$'$ in order to enlarge the point source holes and, therefore, exclude from the analysis additional pixels that can still be affected by their emission. In particular, we mask only those pixels with values $\le$ 0.8.  
 The final point sources mask covers about 1\% of the sky. The masked point sources can be seen in grey in Fig.~\ref{fig:masks}.

\subsection{Total Polarized Intensity Mask}
\label{sub:03,subsec:3}
\begin{figure}[htbp]
\centerline{\includegraphics[scale=.5]{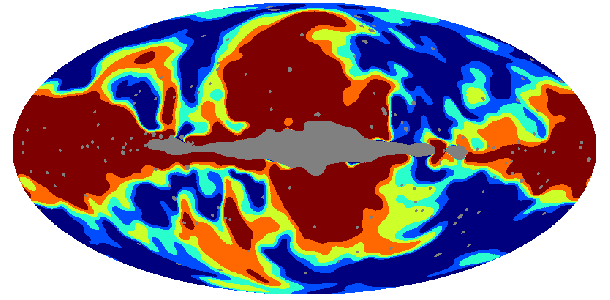}}
\caption{The different regions used to estimate the power spectra are shown. Note that the valid pixels of a given mask are also allowed in all the masks that leave a larger fraction unmasked. In this way, the sky fraction allowed by the 30\% mask is showed in dark red, the valid pixels of the 40\% mask are given by the combination of the dark red and orange regions, the 50\% includes the dark red, orange and yellow areas, the 60\% contains the same regions plus the turquoise one and, finally, the 70\% also includes the light-blue pixels. The 94\% mask only excludes from the analysis the grey region, which corresponds to the combination of the Galactic and point source masks.
}
\label{fig:masks}
\end{figure}

The combination of the previous Galactic and point sources masks defines a preliminary region (of around 6 per cent of the sky) that will be excluded from all of our analysis.
Once these pixels are removed, since for the characterization of the synchrotron emission one should consider regions with a sufficient signal-to-noise ratio, we construct a set of masks that select those areas with the largest polarization signal in the remaining 94 per cent of the sky.
Thus, we mask those pixels below successively lower thresholds of $P$ in the \textit{Planck} 30~GHz polarization map, smoothed to a $5^{\circ}$ resolution (Gaussian beam). The thresholds are chosen such that we select five regions that retain a $f_{sky}$ from 0.3 to 0.7 in steps of 0.1, as shown in Figure \ref{fig:masks}. As one would expect, this procedure tends to mask mostly regions far from the Galactic plane, naturally excluding high latitudes, and leaving unmasked the central regions in both hemispheres. The mask selection process is described in more detail in Appendix \ref{appendix:01}.
In order to have masks with softer boundaries, we smooth them with a Gaussian beam of FWHM=$3^{\circ}$ and exclude all pixels $\leq 0.5$. Furthermore, in the resulting masks, we remove small isolated "holes" and "islands" with radius smaller than $5^{\circ}$, which could otherwise complicate the spectra estimation. In this way, we have a set of 5 masks that, together with the near full sky mask (that removes only the Galactic and point source regions) constitute the basic set of masks for our analysis.
Finally, before calculating the spectra and in order to reduce leakage effects in the power spectra, these six masks are apodized using the “C2” method of \texttt{NaMaster}\footnote{namaster.readthedocs.io} \cite{namaster} with an apodization length of $3^{\circ}$, where pixels are multiplied by a cosine function of their distance to the nearest fully masked pixel. \\
For our main results, we pick as the reference mask the one with $f_{sky}= 0.5$, which is a good compromise between the considered sky fraction and the signal-to-noise ratio, as explained in Appendix \ref{appendix:01}.

%% file: sections/section04.tex
\section{Angular Power Spectra}
\label{sec:04}

In order to characterize the Galactic synchrotron polarization signal, we compute EE and BB auto-spectra and EB cross-spectra on the set of six masks described in the previous section. 
In particular, in order to deal with the presence of masks, which can induce mixing of power between multipoles and between polarization modes, we use the public code \texttt{NaMaster} \cite{namaster}, an advanced implementation of the pseudo-spectrum method to estimate the power spectra on an incomplete sky. Although this type of approach has been shown to be less optimal at large scales than Quadratic-Maximum Likelihood methods, they provide comparable results for the range of scales that we consider in this work, while being significantly faster (see \cite{bilbaoahedo2021eclipse}). \texttt{NaMaster} has already been successfully used in different cosmological applications \cite{Minami_2020,Upham_2021,Azzoni_2021}.

Pseudo-$C_\ell^{EE}$ are estimated with the \texttt{NaMaster} E-purification method, where the so-called \textit{pure} E mode is defined as the field that is orthogonal to all B modes. Similarly, pseudo-$C_\ell^{BB}$ are estimated with the \texttt{NaMaster} B-purification method.
The pseudo-$C_\ell^{EB}$ are computed cross-correlating a \textit{pure} E with a \textit{pure} B field.
Note that at large angular scales (low multipoles) diffuse synchrotron emission dominates, while at higher multipoles ($\ell$ $ \gtrsim$ 250) noise, and possibly extragalactic point sources, can play an important role. Taking this into account, we focus our main analyses in the multipole range 30 $\leq$ $\ell$ $\leq$ 300, binning with $\Delta\ell$ = 10 for multipoles 30 $\leq$ $\ell$ $\leq$ 200 and with $\Delta\ell$ = 20 for multipoles $\ell$ > 200. We discard the lowest multipoles, since pseudo-spectra methods are suboptimal at very large scales on a masked sky. Consistency of the results versus a different choice of the multipole range (10 $\leq$ $\ell$ $\leq$ 400) is discussed in Appendix~\ref{appendix:02,subsection:02}.

We model the EE and BB synchrotron power spectra as a power law parameterized by the index $\alpha$ and amplitude $A$ evaluated on a pivot point $\ell = 80$ 
\begin{equation}
    C_\ell^{XX} = A^{XX} \Big(\frac{\ell}{80}\Big)^{\alpha_{XX}}
\label{fit}
\end{equation}
with $XX = EE, \ BB$. The pivot point at $\ell = 80$ corresponds to the most important scale for measuring the synchrotron contamination to CMB because it is where the maximum of the contribution from cosmological gravitational waves is supposed to be located. We fit the EE and BB power spectrum estimated from the data to the previous model (with a total of 20 degrees of freedom) with a nonlinear least-squares algorithm provided by the \texttt{SciPy} \cite{SciPy} Python packages. Note that before performing the fit, the CMB contribution is subtracted from the data at the spectrum level and this is done by estimating the average CMB power spectra from CMB simulations using the same \texttt{NaMaster} procedure as for the data.
With this procedure, we only subtract the average CMB signal, while the cosmic variance contribution is preserved in the covariance matrices. \\
The EB cross-spectra is simply modelled as a constant
\begin{equation}
    C_\ell^{EB} = A^{EB}.
\label{fitEB}
\end{equation}
In this case we also perform a $\chi^2$ fit, with a total of 21 degrees of freedom. In both fitting processes, we take into account the full covariance matrices computed with simulations. The effects of the instrumental beams and the pixel window function are also considered.

We first fit equations \ref{fit}-\ref{fitEB} independently to the power spectra estimated from \textit{Planck} 30~GHz and from\textit{WMAP} K-band data. Then we fit the EE, BB and EB models to spectra obtained cross-correlating the \textit{WMAP} K-band with the \textit{Planck} 30~GHz maps. The results are presented in the following subsections.
In order to test the robustness of the results, we also fit the same models to the power spectra estimated from the \textit{Planck} PR4 30 GHz \cite{Npipe} data and to the spectra obtained cross-correlating \textit{WMAP} and PR3 in a larger multipole range, results are presented in Appendix \ref{appendix:02}.

\subsection{Planck}
\label{sec:04,subsec:1}
\begin{figure}[htp]
\centerline{\includegraphics[scale=.55]{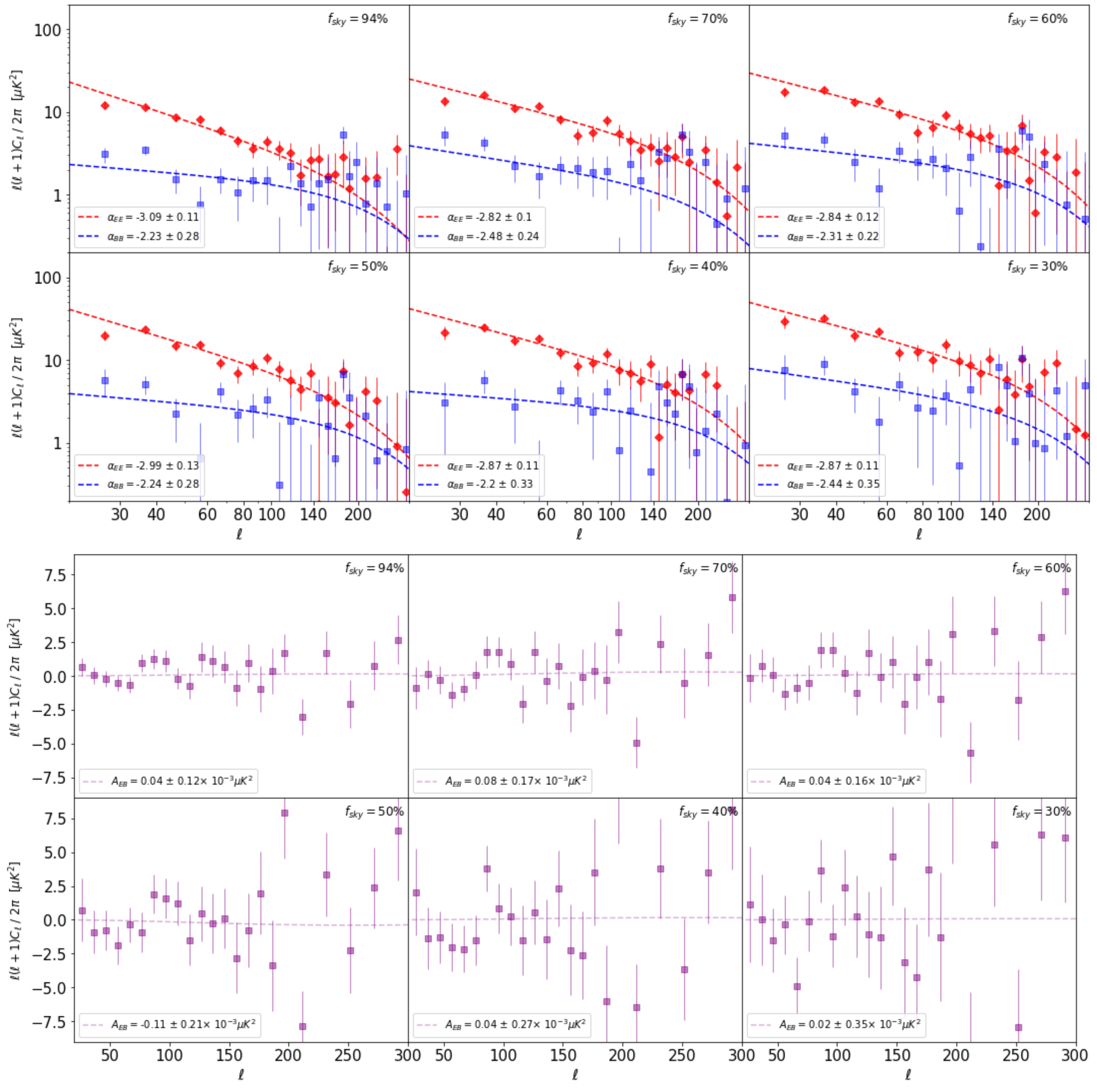}}
\caption{\textit{Planck} results. Top panels: EE (red diamonds), BB (blue squares) pseudo-spectra, bottom panels: EB (purple squares) pseudo-spectra, at the nominal frequency 30 GHz. Spectra are computed cross-correlating the \textit{Planck} 30~GHz half-ring maps, for each of the six sky masks, identified by the unmasked sky fraction. The dashed lines are the best fits to the data points. The indices $\alpha$ (top panels) are the exponent of the fitted power law \ref{fit} and the amplitudes $A_{EB}$ (bottom pannels) are the constant fitted in equation \ref{fitEB}.}
\label{fig:Planck}
\end{figure}

We compute the EE, BB and EB power spectra cross-correlating the half-ring maps of the \textit{Planck}-LFI 30~GHz maps. Covariance matrices are estimated from 100 half-ring simulations. We fit the model in equation \ref{fit} separately for E and B-modes, and for EB we fit a constant (equation \ref{fitEB}).
The power spectra and their corresponding fits are showed in Figure \ref{fig:Planck}, each panel corresponding to a different mask.
Both EE (red diamonds) and BB (blue squares) spectra show a steep decay of the synchrotron amplitude as a function of the scale. 
The goodness of the fit, reported in Table \ref{table:fitPlanck} in terms of $\chi^2$, confirms that the simple power law model describes reasonably well the synchrotron polarization power spectra in most cases. Even so, there are a few $\chi^2$ values, especially for EB, which slightly exceed the expectation for the considered distribution. This seems to be related to the limited number of simulations, which leads to a misestimation of the covariance matrix. Indeed, when repeating the same analysis using PR4 data, for which a larger number of simulations is available, the goodness of fit improves in most cases, especially for the EB fit (see Appendix \ref{appendix:02,subsection:01} for details). The effect of the number of simulations in also further considered in sections  \ref{sec:04,subsec:2} and \ref{sec:04,subsec:3}.

From Table~\ref{table:fitPlanck}, it is seen that, for all the considered masks, there is a systematic difference between the best-fit values for the indices $\alpha_{EE}$ and $\alpha_{BB}$. In particular, for our reference mask ($f_{\rm sky}=0.5$), we find -2.99$\pm$0.13 and -2.24$\pm$0.28, respectively, suggesting a steeper decay of the diffuse synchrotron E-component with respect to the B-component.
By simply combining the errors quadratically, this implies that the two indices are inconsistent at the 2$\sigma$ level. However, there may be correlations between both quantities that are not being taken into account, implying that the combined error may be somewhat underestimated. In addition, the analyses presented in the next sections do not show this behaviour. Therefore, this difference does not seem to be significant.
The B-to-E ratio is computed as the amplitude ratio at the pivot multipole $\ell = 80$, and it turns out to be around 0.27 (slightly varying with the considered sky fraction). This amplitude asymmetry between the two polarization components is confirmed for all the masks, even in the 94\% case, showing that it does not seem to be associated to specific regions but rather to be a feature of the diffusion synchrotron emission in almost the whole sky. \\
The EB cross-spectra is compatible with zero for the whole mask set within 1$\sigma$. We can put an upper constraint on the diffuse synchrotron polarization EB power spectrum, finding it to be smaller than 4.2\% that of the EE spectrum. The constraint is computed from the 2$\sigma$ error bar of the amplitude ratio $A^{EB}/A^{EE}$ of the reference mask in Table \ref{table:fitPlanck}, with the ratio evaluated at the pivot multipole $\ell = 80$. As far as we know, this is the first direct constraint on the EB cross-correlation of the diffuse synchrotron emission. \\

We can compare our results with those found by the \textit{Planck} Collaboration in 2018 using the Commander and SMICA component separation methods \cite{akrami2020planck}. They estimated the spectra from a different region, which considers intermediate and high Galactic latitudes, thus finding smaller amplitudes than in our analysis, $A^{EE}=2.3\pm0.1$ ($2.4\pm0.1$) $\mu K^2$ for Commander (SMICA), but very compatible\footnote{Along this work, when referring to compatibility between two values (either obtained here or in relation to previous results), unless otherwise stated, we mean that the difference of the two central values is less than $2\sigma$. In this case we take $\sigma = \sqrt{\sigma_1^2 + \sigma_2^2}$ with $\sigma_i$ the errors of the compared values. Note that this combined error is just an approximation, since it does not take into account possible correlations between the two considered quantities.}
EE power spectrum index $\alpha_{EE}=-2.84\pm0.05$ ($-2.88\pm0.04$). The value found with Commander (SMICA) for $\alpha_{BB}=-2.76\pm0.09$ ($-2.75\pm0.07$) is slightly different from ours, even if compatible at 2$\sigma$. This apparent discrepancy can be explained by the very different procedure used in that work to extract the foreground signal (through a component separation method that takes into account all frequency channels), the use of different sky regions as well as a different power spectrum estimation method. Those differences are expected to have a larger impact on the BB spectrum, rather than EE, because of the lower signal-to-noise.
The B-to-E ratio, around 0.34, is very compatible with the value we find for our 94\% mask, that is 0.33$\pm$0.06, which is the most similar mask to that used in the \textit{Planck} work. Nevertheless, the B-to-E ratio compatibility holds also for our 30\% mask, that is 0.29$\pm$0.04, which instead left unmasked mostly low latitudes. \\
Comparing our results with the ones found in 2018 with \textit{S-PASS} \cite{NicolettaSych} at frequency 2.3 GHz, we see some differences. In particular, the BB synchrotron spectra is found to have a steeper decay, with $\alpha_{BB}$ around -3, and the B-to-E ratio is about 0.5 (although the specific values of these parameters vary strongly with the considered latitudes). These differences could be explained by the different frequency of observation of both experiments and, to a lesser extent, by the different regions considered for the analysis. Indeed, at low frequencies we expect other physical effects to take place, such as the Faraday rotation, depolarizing the synchrotron emission. 

\begin{table}[htp]
\centering
\scalebox{0.78}{
\begin{tabular}{|c|c|c|c|c|c|c|}
\hline
$f_{sky}$ & \textbf{94\%} &  \textbf{70\%} & \textbf{60\%} & \textbf{50\%} & \textbf{40\%} & \textbf{30\%} \\ [0.5ex]
 \hline
   \rule{0pt}{3ex}
 $\alpha_{EE}$  &  -3.09 $\pm$ 0.11 &   -2.82 $\pm$ 0.10 &  -2.84 $\pm$ 0.12 &  -2.99 $\pm$ 0.13 &  -2.87 $\pm$ 0.11 &  -2.87 $\pm$ 0.11 \\
$\alpha_{BB}$  &  -2.22 $\pm$ 0.28 &  -2.48 $\pm$ 0.24 &  -2.31 $\pm$ 0.22 &  -2.24 $\pm$ 0.28 &   -2.20 $\pm$ 0.33 &  -2.44 $\pm$ 0.35 \\ [1ex]
\rule{0pt}{3ex}
 $A^{EE}$ $[10^{-3}\mu K]$ &   4.86 $\pm$ 0.28 &   7.69 $\pm$ 0.44 &   8.87 $\pm$ 0.52 &  10.01 $\pm$ 0.62 &  12.07 $\pm$ 0.61 &   14.4 $\pm$ 0.69 \\
$A^{BB}$ $[10^{-3}\mu K]$  &   1.62 $\pm$ 0.27 &   1.92 $\pm$ 0.29 &    2.61 $\pm$ 0.30 &   2.72 $\pm$ 0.37 &   3.04 $\pm$ 0.45 &   4.14 $\pm$ 0.59 \\
$A^{BB}/A^{EE}$ &   0.33 $\pm$ 0.06 &   0.25 $\pm$ 0.04 &   0.29 $\pm$ 0.04 &   0.27 $\pm$ 0.04 &   0.25 $\pm$ 0.04 &   0.29 $\pm$ 0.04 \\ [1ex]
 \rule{0pt}{2.5ex}
 $\chi_{EE}^2$ (20 dof) &        19.3 &        22.0 &        25.6 &        29.8 &        21.9 &         19.5 \\
$\chi_{BB}^2$ (20 dof) &        35.3 &        20.3 &        20.6 &         22.1 &        21.7 &        25.2 \\ [1ex] 
 \hline
 \rule{0pt}{3ex}
 $A^{EB}$ $[10^{-3}\mu K]$ &    0.04 $\pm$ 0.12 &   0.08 $\pm$ 0.17 &    0.04 $\pm$ 0.16 &    -0.11 $\pm$ 0.21 &    0.04 $\pm$ 0.27 &    0.02 $\pm$ 0.35 \\
 $A^{EB}/A^{EE}$ &  0.008 $\pm$ 0.024 &  0.010 $\pm$ 0.022 &  0.005 $\pm$ 0.018 &  -0.011 $\pm$ 0.021 &  0.004 $\pm$ 0.022 &  0.002 $\pm$ 0.024 \\ [1ex]
 \rule{0pt}{2.5ex}
 $\chi_{EB}^2 $ (21 dof)  &          30.4 &         34.5 &          27.8 &           37.1 &          41.1 &          32.2 \\ [1ex]
 \hline
\end{tabular}}
\caption{\textit{Planck} results. Best-fit parameters with 1$\sigma$ errors and $\chi^2$ of the power-law in equation \ref{fit} for EE and BB, and of the constant baseline in \ref{fitEB} for EB. Spectra are computed cross-correlating the \textit{Planck} 30~GHz half-ring maps, for each of the six sky masks described in section \ref{sec:03}.}
\label{table:fitPlanck}
\end{table}

\subsection{WMAP}
\label{sec:04,subsec:2}
\begin{figure}[htp]
\centerline{\includegraphics[scale=.55]{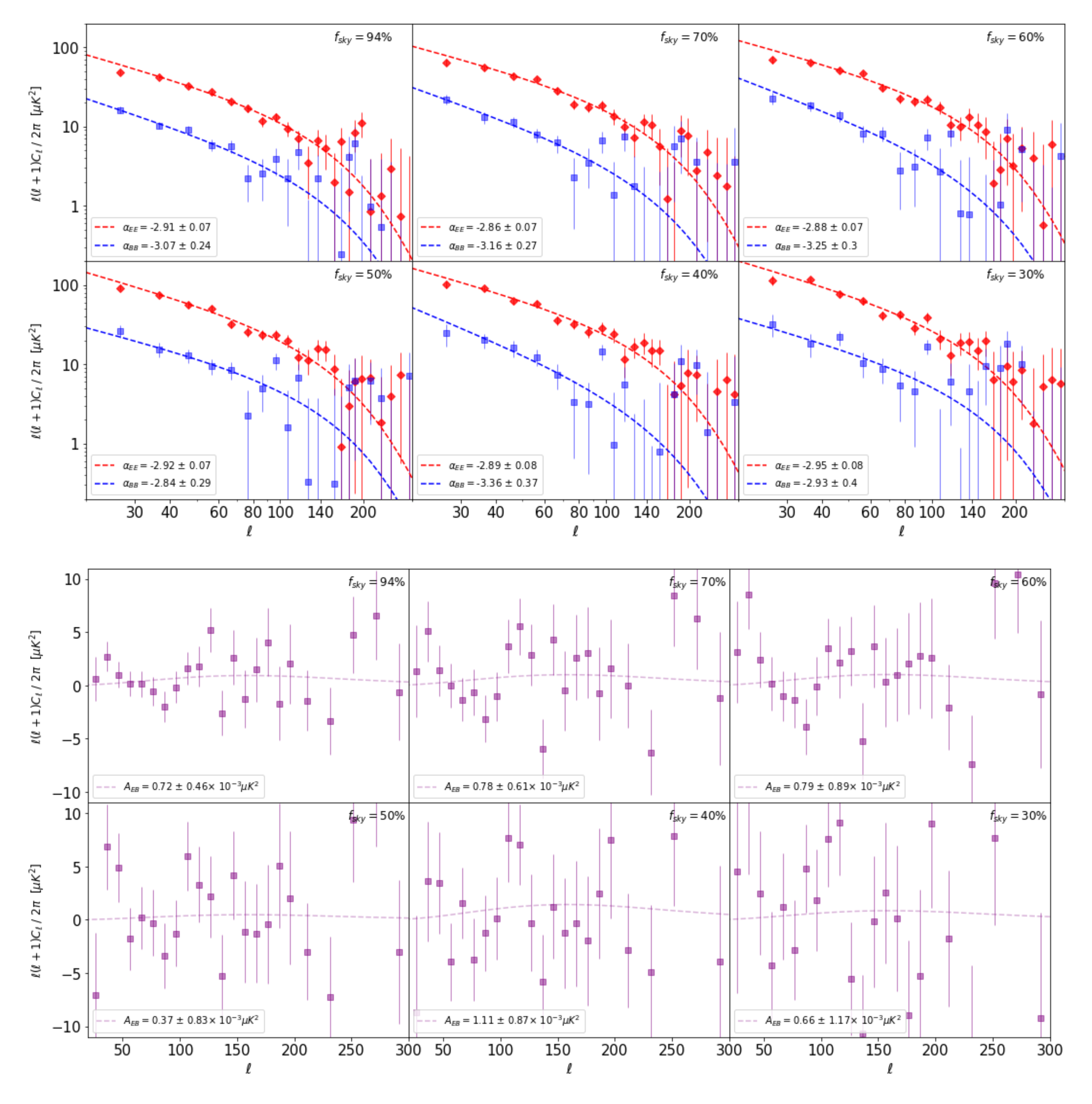}}
\caption{\textit{WMAP} results. Top: EE (red diamonds), BB (blue squares) pseudo-spectra, bottom: EB (purple squares) pseudo-spectra, at the nominal frequency 23 GHz. Spectra are computed cross-correlating the co-added \textit{WMAP} K-band year maps, for each of the six sky masks, identified by the unmasked sky fraction. The dashed lines are the best fits to the data points. The indices $\alpha$ (top) are the exponent of the fitted power law \ref{fit} and the amplitudes $A_{EB}$ (bottom) are the constant fitted in equation \ref{fitEB}.}
\label{fig:WMAP}
\end{figure}

We compute the EE, BB and EB power spectra for the \textit{WMAP} K-band cross-correlating the co-added year maps from 1 to 4 with the co-added year maps from 5 to 9. Covariance matrices are estimated from 300 simulations per year-split\footnote{Note that in the case of \textit{Planck} half-ring maps, the number of simulations were limited to 100, as provided by the \textit{Planck} collaboration. For \textit{WMAP}, we tested the robustness of the results versus the number of simulations. We found that results were stable with 300 simulations, improving in particular the error in $\alpha_{EE}$ with respect to the use of a lower number of simulations.}.
We fit the model in equation \ref{fit} separately for E and B-modes, and for EB we fit the model of equation \ref{fitEB}. The power spectra and the corresponding fits are showed in Figure~\ref{fig:WMAP} while the best-fits parameters are given in Table~\ref{table:fitWMAP}.

As for \textit{Planck}, EE and BB spectra show both a steep decay of the synchrotron amplitude as a function of scale.
For the reference mask ($f_{sky} = 50\%$), we find $\alpha_{EE}=-2.92 \pm 0.07$ and $\alpha_{BB}=-2.84 \pm 0.29,$ very compatible between them. This kind of consistency holds for most of the mask set, supporting the hypothesis that the two synchrotron polarization components decay as a function of multipoles with the same ratio. The estimated B-to-E ratio ranges from $0.23 \pm 0.03$ (for the 50 and 94\% masks) to $0.17 \pm 0.03$ (when considering 40\% of the sky).
This amplitude ratio between the two polarization components suggests an even stronger asymmetry than the one found for the \textit{Planck} 30~GHz channel.

The EB cross-correlation is compatible with zero at 1$\sigma$ in the reference mask and within 2$\sigma$ for the whole mask set. This is a further confirmation of the hypothesis of null cross-component in the diffuse synchrotron polarization. The EB/EE ratio evaluated at the pivot multipole $\ell = 80$ , 
provides an upper limit to the EB amplitude $ A_{EB}\le 0.044~A_{EE}$ at $2\sigma$. 

\begin{figure}[ht]
\centerline{\includegraphics[scale=.41]{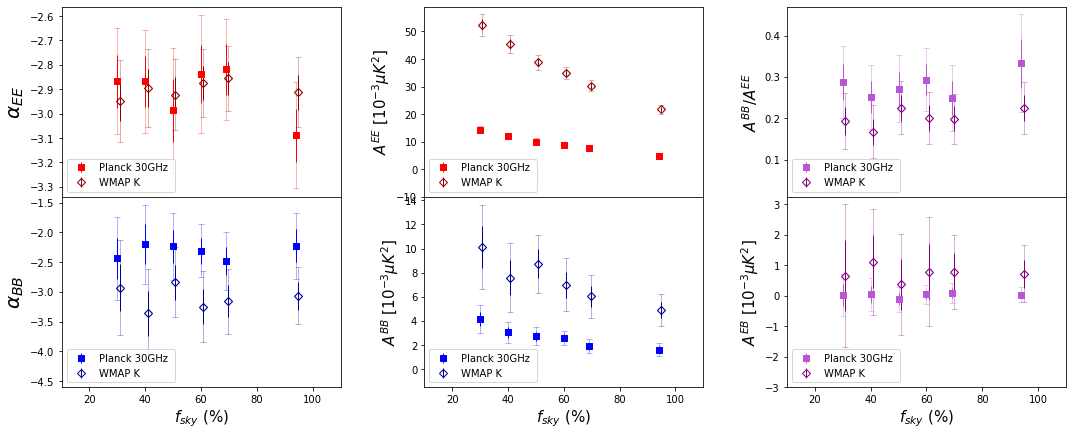}}
\caption{Comparison between the best-fit parameters found for \textit{Planck} 30~GHz (filled squares) and \textit{WMAP} K-band (blank diamonds) to the models of equations \ref{fit} and \ref{fitEB}. $1\sigma$ and $2\sigma$ errors are showed with thick and thin lines, respectively. To allow for a better visualization, \textit{WMAP} values are slightly shifted in the x-axis.}
\label{fig:Planck_WMAP}
\end{figure}

The model of the synchrotron polarization spectra derived from \textit{WMAP}, for most of the masks, is consistent with the power law model with null EB cross-correlation that we obtained for \textit{Planck}, as showed in Figure \ref{fig:Planck_WMAP}, as well as with the Commander and SMICA models \cite{akrami2020planck}.
However, \textit{WMAP} data suggests a slightly smaller B-to-E ratio, where the E-component is about 4.3 times larger than the B-component. This is again also different from the \textit{S-PASS} results \cite{NicolettaSych}, but, as previously mentioned, this could be explained by the different frequency and, to a lesser extent, by the different regions observed by both experiments.

\begin{table}[htp]
\centering
\scalebox{0.78}{
\begin{tabular}{|c|c|c|c|c|c|c|}
\hline
$f_{sky}$ & \textbf{94\%} &  \textbf{70\%} & \textbf{60\%} & \textbf{50\%} & \textbf{40\%} & \textbf{30\%} \\ [0.5ex] 
\hline
\rule{0pt}{3ex}
$\alpha_{EE}$ &  -2.91 $\pm$ 0.07 &  -2.86 $\pm$ 0.07 &  -2.88 $\pm$ 0.07 &  -2.92 $\pm$ 0.07 &  -2.89 $\pm$ 0.08 &  -2.95 $\pm$ 0.08 \\
$\alpha_{BB}$ &  -3.07 $\pm$ 0.24 &  -3.16 $\pm$ 0.27 &   -3.25 $\pm$ 0.30 &  -2.84 $\pm$ 0.29 &  -3.36 $\pm$ 0.37 &   -2.93 $\pm$ 0.40 \\ [1ex]
\rule{0pt}{3ex}
$A^{EE}$ $[10^{-3}\mu K]$ &  21.83 $\pm$ 0.83 &  30.31 $\pm$ 1.02 &  34.87 $\pm$ 1.15 &  38.75 $\pm$ 1.29 &   45.45 $\pm$ 1.70 &  52.38 $\pm$ 1.99 \\
$A^{BB}$ $[10^{-3}\mu K]$ &   4.93 $\pm$ 0.66 &   6.03 $\pm$ 0.89 &    7.00 $\pm$ 1.09 &    8.73 $\pm$ 1.20 &    7.60 $\pm$ 1.43 &  10.16 $\pm$ 1.75 \\
$A^{BB}/A^{EE}$ &   0.23 $\pm$ 0.03 &    0.20 $\pm$ 0.03 &    0.20 $\pm$ 0.03 &   0.23 $\pm$ 0.03 &   0.17 $\pm$ 0.03 &   0.19 $\pm$ 0.03 \\ [1ex]
\rule{0pt}{2.5ex}
$\chi_{EE}^2$ (20 dof) &        21.4 &        16.3 &        17.4 &        17.1 &        15.8 &        16.0 \\
$\chi_{BB}^2$ (20 dof) &        23.0 &        23.0 &        24.8 &        22.9 &        28.7 &        29.4 \\ [1ex] 
\hline
\rule{0pt}{3ex}
$A^{EB}$ $[10^{-3}º\mu K]$ &    0.72 $\pm$ 0.46 &   0.78 $\pm$ 0.61 &    0.79 $\pm$ 0.89 &   0.37 $\pm$ 0.83 &    1.11 $\pm$ 0.87 &    0.66 $\pm$ 1.17 \\
$A^{EB}/A^{EE}$ &  0.033 $\pm$ 0.021 &  0.026 $\pm$ 0.020 &  0.023 $\pm$ 0.026 &  0.010 $\pm$ 0.022 &  0.024 $\pm$ 0.019 &  0.013 $\pm$ 0.022 \\  [1ex]
\rule{0pt}{2.5ex}
$\chi_{EB}^2$ (21 dof) &          23.0 &         24.6 &          35.6 &         27.3 &          22.9 &          31.3 \\ [1ex]
\hline
\end{tabular}}
\caption{\textit{WMAP} results. Best-fit parameters with 1$\sigma$ errors and $\chi^2$ of the power-law in equation \ref{fit} to EE and BB, and of the constant baseline in \ref{fitEB} to EB. Spectra are computed cross-correlating the co-added \textit{WMAP} K-band years maps, for each of the six sky masks described in section \ref{sec:03}.}
\label{table:fitWMAP}
\end{table}

\subsection{WMAP-Planck cross spectra}
\label{sec:04,subsec:3}
From the previous \textit{WMAP} and \textit{Planck} individual analysis, we can conclude that our characterization of the synchrotron polarization power spectra, provided the Galactic centre and bright point sources are properly masked, holds reasonably well in the frequency range 23-30 GHz and is supported by two independent experiments. 
Therefore, we can improve the estimation of the model parameters by cross-correlating the data of the two experiments, increasing the signal-to-noise while reducing the effect of possible systematic errors.
Differently from the previous analyses, we use the full-mission \textit{Planck} 30 GHz and the co-added 9 years \textit{WMAP} K-band maps instead of data splits. 
Since we are cross-correlating two independent experiments, instrumental noise will cancel and possible systematics will be reduced even when using the full data maps. In addition, we can benefit of the larger number of simulations available, that yields to smaller errors on the estimated parameters. 

\begin{figure}[ht]
\centerline{\includegraphics[scale=.40]{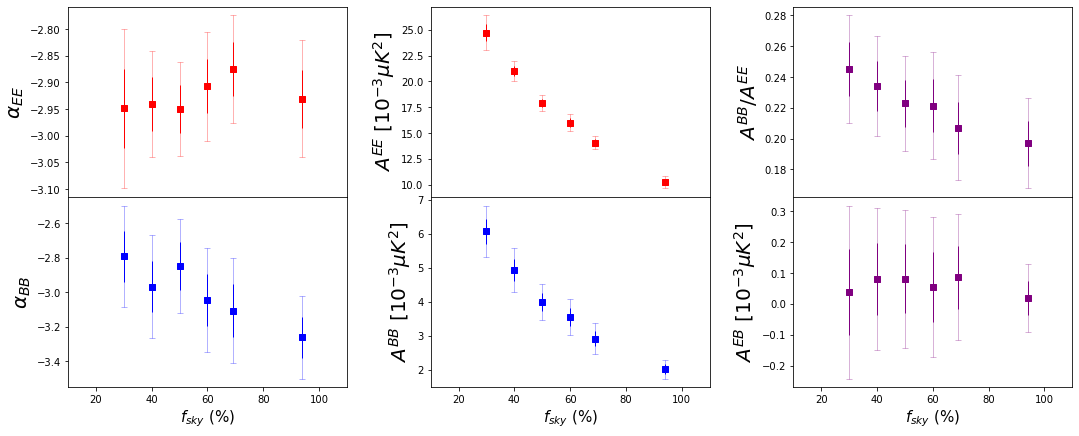}}
\caption{\textit{Planck}-\textit{WMAP} results. Best-fit parameters to the models of equations \ref{fit} and \ref{fitEB} computed on the cross-spectra of \textit{WMAP} K-band and \textit{Planck} 30GHz. 1$\sigma$ and $2\sigma$ errors are showed with thick and thin lines, respectively.}
\label{fig:Planck_WMAP_cross}
\end{figure}

We generated 600 spectra cross-correlating Planck and WMAP simulations to estimate the covariance matrices. Being the \textit{Planck} noise simulations limited to 300, we used each of these noise simulations twice\footnote{A similar procedure has been followed to increase the number of simulations in \cite{Planck_IS}.}. Even if the simulations are not completely independent because of the limited number of the Planck noise simulations, the other two components (CMB and WMAP noise) are still fully independent, giving a better statistics which improves the estimation of the covariance matrix.
Although for the sake of brevity we omit the results for the \textit{Planck}-\textit{WMAP} cross-spectra derived from the split data, they are fully compatible with the results presented in this section.
Fig.~\ref{fig:Planck_WMAP_cross} shows the EE, BB and EB power spectra and the best fits obtained from the \textit{WMAP}-\textit{Planck} cross-correlation analysis, Table~\ref{table:fitCross} provides the best-fit parameters and the $\chi^2$ values. 

For our reference mask ($f_{sky}=50\%$), the EE and BB power spectra of the diffuse synchrotron emission show a steep decay as a function of multipoles with consistent power spectrum indices $\alpha_{EE} = -2.95 \pm0.04$  and $\alpha_{BB} = -2.85 \pm 0.14$. The goodness of the fits supports the validity of the power law model and the hypothesis of compatibility between the steepness of the two polarization components. Considering the results for the full set of masks shown in Figure \ref{fig:Planck_WMAP_cross}, $\alpha_{EE}$ is very stable and compatible with the nearly-full sky case (94\%). Instead, the BB power law shows a slight tendency to steeper values when including high latitudes. The B-to-E ratio for the reference mask is found to be $0.22 \pm 0.02$ and it ranges from 0.20 in the 94\% mask to 0.25 for the 30\% mask, which shows again how the foreground E-component dominates the polarization emission over the B-component at low frequency. \\
We find that the difference between the two polarization components, both in amplitude and steepness, is larger when considering the nearly full-sky case, whereas it tends to decrease when considering mainly regions closer to the Galactic plane (where the signal-to-noise is higher). In any case, the results computed for the complete set of masks are consistent at 2$\sigma$, except for the 94\% mask (where consistency holds at 2.5$\sigma$). A more precise characterization of the EE and BB power spectra with latitude would require maps with better sensitivity in polarization.

The EB cross-spectra is consistent with zero at 1$\sigma$ for the whole mask set, therefore, the hypothesis of null EB cross-correlation holds for the diffuse synchrotron emission even when considering the nearly full-sky case. 
The spectra provided by the cross-correlation of \textit{Planck} and \textit{WMAP} gives the most stringent upper limit to the EB amplitude, found to be $\le$ 1.2\% (2$\sigma$) of the EE amplitude, computed at multipole $\ell = 80$. \\
As anticipated in section \ref{sec:04,subsec:1}, we point out that the covariance matrices used in the fit for EB could be underestimated because of the limited number of simulations, leading to some larger $\chi^2$ values. We tested the dependency of $\chi^2$ on the number of simulations using samples of different sizes, finding that a larger number of simulations lead to more stable results and to lower values of $\chi^2$.

These results point out that the two most important polarised foregrounds, thermal dust and synchrotron, present some differences at spectra level. At high frequency the thermal dust emission shows a power law decay with power spectrum index $\alpha_d\approx$ -2.5 \cite{planckDust}, less steep than the power spectrum index $\alpha_s\approx$ -2.9 that we find for synchrotron. Moreover, the B/E asymmetry for the synchrotron emission, around 0.22, is stronger than the asymmetry found for dust, around 0.5. The observed EB/EE power ratio, for both thermal dust and synchrotron, is smaller than about 0.03.

The results presented in this section for the cross-experiment analysis are, in general, compatible with the single-experiment analysis of \textit{Planck} and \textit{WMAP}. Moreover, the steepness of the power law model which describes the EE and BB power spectra are compatible with the \textit{Planck} component separation results presented in 2018 \cite{akrami2020planck}. Nevertheless, the B-to-E ratio we find in the nearly full sky case is slightly smaller than the ratio found by \textit{Planck} in the intermediate and high latitudes. As discussed in the previous sections, several differences appear when comparing our results with the model found with \textit{S-PASS} \cite{NicolettaSych}, mainly due to the fact that they analyze maps at a lower frequency, that is 2.3 GHz, and only at high Galactic latitudes. 
In particular, we find a less steep and compatible decay for both EE and BB and a B-to-E ratio of about 0.22, where instead with \textit{S-PASS} it is observed that at Galactic latitudes $|b|>30^\circ$ the mean value of the decay index is $\alpha\simeq-3.15$ and with a B-to-E ratio $\simeq0.5$ for $|b| > 35^\circ$. 

From our analysis and the comparison with the two mentioned works, we can conclude that our characterization of the synchrotron polarization power spectra can be extended to high Galactic latitude, thus to the full sky after removing the brightest regions, in the frequency range 23-30 GHz. Instead, we cannot exclude that at different frequencies, in particular smaller, the synchrotron polarization power spectra could deviate from our characterization due to some physical effects.

In Appendix \ref{appendix:02} some robustness tests are presented, which confirm the results obtained in 
section~\ref{sec:04}. In particular, in section \ref{appendix:02,subsection:01} we repeat the same analysis cross-correlating the A/B detector splits of the 2020 \textit{Planck} \texttt{NPIPE} release (PR4). In section \ref{appendix:02,subsection:02}, we test the same power law model in a larger multipole range (10 $\leq$ $\ell$ $\leq$ 400). Finally, in Appendix \ref{appendix:03}, we also analyse and test the model independently in the Northern and Southern hemisphere, finding consistency with the results presented in this section. \\

\begin{figure}[htp]
\centerline{\includegraphics[scale=.55]{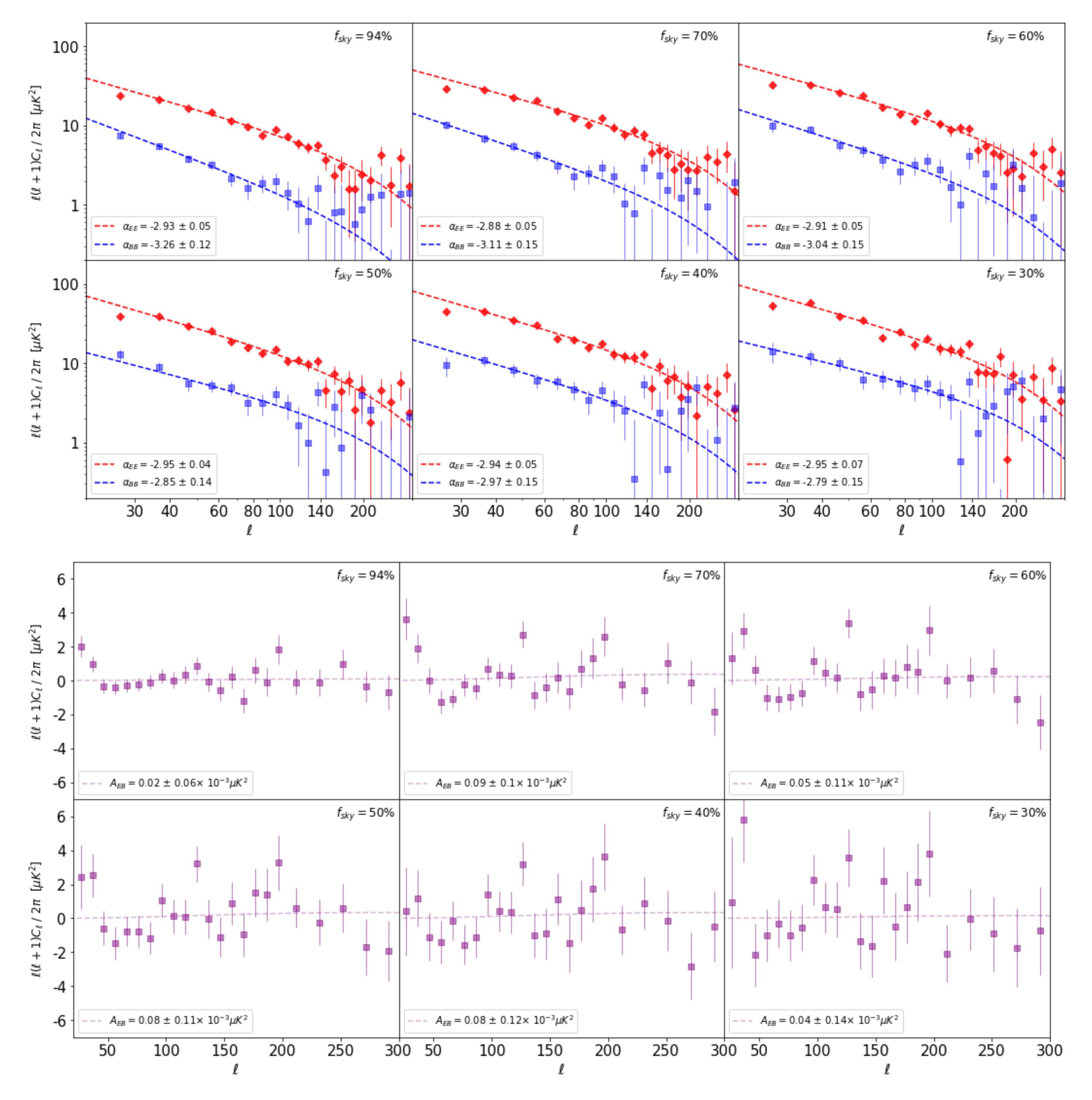}}
\caption{\textit{Planck}-\textit{WMAP} results. Top: EE (red diamonds), BB (blue squares) pseudo-spectra, bottom: EB (purple squares) pseudo-spectra. Spectra are computed cross-correlating the co-added 9 year \textit{WMAP} K-band maps and the full-mission \textit{Planck} 30GHz maps, for each of the six sky masks, identified by the unmasked sky fraction. The dashed lines are the best fits to the data points. The indices $\alpha$ (top) are the exponent of the fitted power law \ref{fit} and the amplitudes $A_{EB}$ (bottom) are the constant fitted in equation \ref{fitEB}.}
\label{fig:Cross}
\end{figure}

\begin{table}[ht]
\centering
\scalebox{0.78}{
\begin{tabular}{|c|c|c|c|c|c|c|}
\hline
$f_{sky}$ & \textbf{94\%} &  \textbf{70\%} & \textbf{60\%} & \textbf{50\%} & \textbf{40\%} & \textbf{30\%} \\ [0.5ex]
 \hline
   \rule{0pt}{3ex}
 $\alpha_{EE}$  &  -2.93 $\pm$ 0.05 &  -2.88 $\pm$ 0.05 &  -2.91 $\pm$ 0.05 &  -2.95 $\pm$ 0.04 &  -2.94 $\pm$ 0.05 &  -2.95 $\pm$ 0.07 \\
$\alpha_{BB}$  &  -3.26 $\pm$ 0.12 &  -3.11 $\pm$ 0.15 &  -3.04 $\pm$ 0.15 &  -2.85 $\pm$ 0.14 &  -2.97 $\pm$ 0.15 &  -2.79 $\pm$ 0.15 \\ [1ex]

  \rule{0pt}{3ex}
 $A^{EE}$ $[10^{-3}\mu K]$ &   10.26 $\pm$ 0.30 &  14.09 $\pm$ 0.33 &   16.02 $\pm$ 0.40 &  17.88 $\pm$ 0.39 &   21.02 $\pm$ 0.50 &  24.72 $\pm$ 0.83 \\
$A^{BB}$ $[10^{-3}\mu K]$  &   2.02 $\pm$ 0.14 &   2.92 $\pm$ 0.23 &   3.55 $\pm$ 0.26 &   3.99 $\pm$ 0.26 &   4.92 $\pm$ 0.32 &   6.06 $\pm$ 0.38 \\
$A^{BB}/A^{EE}$ &    0.20 $\pm$ 0.01 &   0.21 $\pm$ 0.02 &   0.22 $\pm$ 0.02 &   0.22 $\pm$ 0.02 &   0.23 $\pm$ 0.02 &   0.25 $\pm$ 0.02 \\ [1ex]
 \rule{0pt}{2.5ex}
 $\chi_{EE}^2$ (20 dof) &        28.3 &        19.9 &        20.2 &        13.9 &        15.9 &        30.2 \\
$\chi_{BB}^2$ (20 dof) &          12.7 &        17.7 &        18.5 &        14.5 &        16.2 &        15.1 \\ [1ex] 
 \hline
 \rule{0pt}{3ex}
 $A^{EB}$ $[10^{-3}\mu K]$ &    0.02 $\pm$ 0.06 &     0.09 $\pm$ 0.10 &    0.05 $\pm$ 0.11 &    0.08 $\pm$ 0.11 &    0.08 $\pm$ 0.12 &    0.04 $\pm$ 0.14 \\
 $A^{EB}/A^{EE}$ &  0.002 $\pm$ 0.005 &  0.006 $\pm$ 0.007 &  0.003 $\pm$ 0.007 &  0.005 $\pm$ 0.006 &  0.004 $\pm$ 0.005 &  0.002 $\pm$ 0.006 \\ [1ex]
 \rule{0pt}{2.5ex}
 $\chi_{EB}^2 $ (21 dof) &          23.0 &          37.1 &          36.4 &          29.5 &          24.6 &          23.6 \\ [1ex]
 \hline
\end{tabular}}
\caption{\textit{Planck}-\textit{WMAP} results. Best-fit parameters, 1$\sigma$ errors and $\chi^2$ values for the power-law in equation \ref{fit} for EE and BB, and for the constant baseline in \ref{fitEB} for EB. Power spectra are computed by cross-correlating the co-added 9 year \textit{WMAP} K-band maps and the full-mission \textit{Planck} 30~GHz maps, for each of the six sky masks described in section \ref{sec:03}.}
\label{table:fitCross}
\end{table}

%% file: sections/section05.tex
\section{Spectral Index}
\label{sec:05}
In the previous section we focused our analysis on the characterization of the polarization power spectra of \textit{Planck} and \textit{WMAP} at low frequency. However, since we have observations at different frequencies, nominally at 23 and 30 GHz, we can also get insights of the behaviour of the diffuse synchrotron polarization with frequency.

\subsection{Methodology}

The synchrotron spectral energy distribution (SED), both in temperature and polarization, is generally described for each pixel by a power law\footnote{
More complex models are also considered in the literature, such as the one including a curvature parameter. However, given the sensitivity of \textit{Planck} and \textit{WMAP} data, we restrict our analysis to the simple power law case. Future low frequency data, such as \textit{QUIJOTE}, would help to discriminate between these two types of models.}
\begin{equation}
S= S_0~\left( \frac{\nu}{\nu_0}\right)^{\beta}
\label{eq:SED}
\end{equation}
where $S_0$ is the foreground amplitude of a particular pixel at the pivot frequency $\nu_0$ and $\beta$ is the energy spectral index that we assume spatially constant for simplicity. The modelling of the synchrotron SED in polarization as well as the knowledge of the spectral index $\beta$, are essential to test and perform component separation in current and future CMB polarization studies. \\
From equation \ref{eq:SED}, we can get the relationship between the amplitudes of the power spectra of the \textit{WMAP} K-band and the \textit{Planck} 30~GHz maps for the two polarization components E and B
\begin{equation}
(A^{XX})^{WMAP} = (A^{XX})^{Planck} ~\left( \frac{\nu^{WMAP}}{\nu^{Planck}}\right)^{2\beta_{XX}}
\label{AmpWMAP}
\end{equation}
with $\nu^{WMAP}$ = 23 GHz, $\nu^{Planck}$ = 28.4 GHz and $XX = EE, \ BB$. \\
Combining equations \ref{AmpWMAP} and \ref{fit}, we get a system of equations that relate the energy spectral index $\beta$ and the power spectrum index $\alpha$ for each of the polarization components 
\begin{equation}
\begin{cases}
  (C_\ell^{XX})^{Planck} = (A^{XX})^{Planck} \left( \frac{\ell}{80}\right)^{\alpha_{XX}} \\
  (C_\ell^{XX})^{WMAP} = (A^{XX})^{Planck} \left( \frac{\ell}{80}\right)^{\alpha_{XX}} \left( \frac{\nu^{WMAP}}{\nu^{Planck}} \right)^{2\beta_{XX}}
\end{cases}
\label{eq:fit_beta}
\end{equation}
with $XX = EE, \ BB$, and $(A^{XX})^{Planck}$ refers to the polarization power spectrum amplitude at $\ell = 80$ for the \textit{Planck} lowest channel, centered at 28.4 GHz. Before performing the fit, in order to account for the effect introduced by the instrumental bandpasses, we corrected the amplitudes in equations \ref{eq:fit_beta} multiplying them by the colour correction coefficients, following the same procedure and using the same coefficients as those given in \cite{planckDust}.
We perform a $\chi^2$ fit to the system of equations \ref{eq:fit_beta} for the EE and BB auto-spectra with 41 degrees of freedom, keeping the amplitude $A$, the power spectrum index $\alpha$ and the energy spectral index $\beta$ as free parameters. The first two parameters have been widely tested in the previous section, providing a robust test to validate the results presented in this section.
In particular, for the analysis, we use the EE and BB power spectra obtained in section~\ref{sec:04,subsec:1} for \textit{Planck} and section \ref{sec:04,subsec:2} for \textit{WMAP}.

\subsection{Results}
Results are reported in Table \ref{table:Cross_beta}. For our reference mask (50\%), the spectral indices $\beta_{EE}$ and $\beta_{BB}$ are very consistent, with values of -3.00$\pm$0.10 and -3.05$\pm$0.36 respectively. These values are consistent with the spectral index found for intensity by previous \textit{Planck}/\textit{WMAP} analyses \cite{davies2006determination,ade2015joint}, supporting the model of a steep decay due to radiative losses, which cause spectral ageing \cite{dickinson2016cmb}. Moreover, our values are consistent with results found  by \textit{S-PASS} \cite{NicolettaSych} in polarization and by the \textit{Planck} component separation methods \cite{akrami2020planck}. 
We note, however, that several $\chi^2$ values corresponding to the B component exceed the expected values for the considered degrees of freedom. As already discussed in section \ref{sec:04}, this seems to be related to the limited number of \textit{Planck} simulations, having probably a larger impact on BB due to the smaller signal-to-noise.

\begin{figure}[h]
\centerline{\includegraphics[scale=.41]{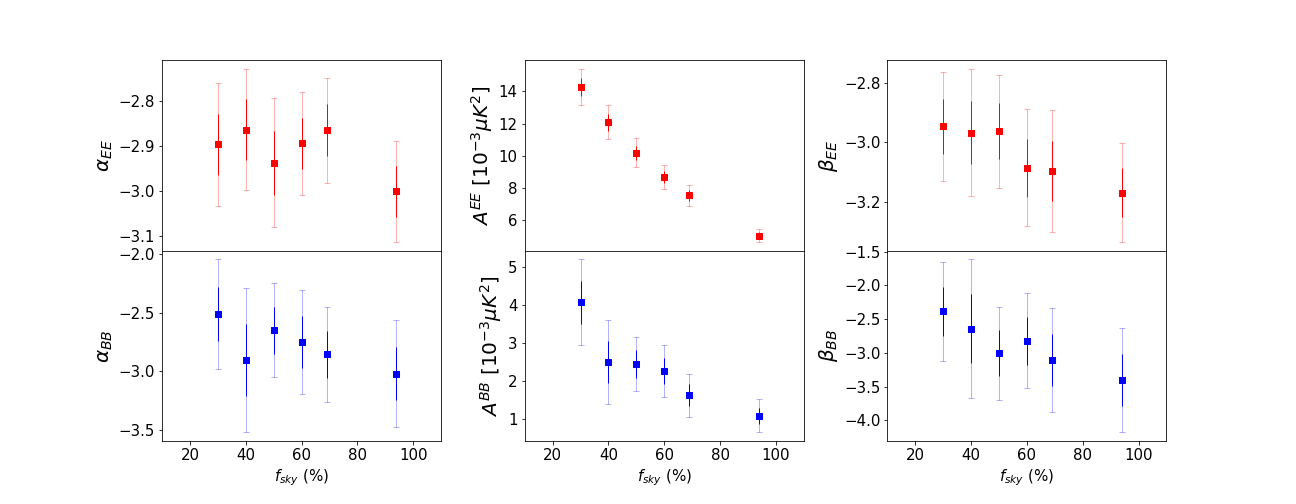}}
\caption{ Best fit parameters to the system of equations of \ref{eq:fit_beta} computed from the \textit{WMAP} K-band and \textit{Planck} 30~GHz data. 1$\sigma$ and 2$\sigma$ errors are showed with thick and thin lines, respectively. $A^{XX}$ refers to the polarization power spectrum amplitude at $\ell = 80$ for the \textit{Planck} lowest channel.}
\label{fig:Cross_beta}
\end{figure}

Figure \ref{fig:Cross_beta} shows the best-fit parameters for each of the considered masks for both the E- and B-mode components. It is interesting to point out that the power spectral indices tend to move towards steeper values when considering larger sky fractions, i.e., when including higher Galactic latitudes in the analysis. In particular, $\beta_{EE}$ ranges from -2.98 to -3.22 while $\beta_{BB}$ expands a wider range, from -2.39 to -3.48, but with larger uncertainties. This kind of steepening has been observed in other previous works \cite{NicolettaSych,FuskelandSynch}, showing that the spectral index gradually steepens from $\beta$ $\simeq$ -2.8 to $\beta$ $\simeq$ -3.3 when including higher Galactic latitudes.

 We also note that the power spectrum indices $\alpha_{EE}$, $\alpha_{BB}$ and the B-to-E ratio are compatible with the results presented in section \ref{sec:04,subsec:3}, supporting the robustness of the analysis presented in this section. \\

\begin{table}[h]
\centering
\scalebox{0.81}{
\begin{tabular}{|c|c|c|c|c|c|c|}
\hline
$f_{sky}$ & \textbf{94\%} &  \textbf{70\%} & \textbf{60\%} & \textbf{50\%} & \textbf{40\%} & \textbf{30\%} \\ [0.5ex] 
 \hline
 \rule{0pt}{3ex}
 $\alpha_{EE}$ &   -3.00 ± 0.06 &  -2.87 ± 0.06 &  -2.89 ± 0.06 &  -2.94 ± 0.07 &  -2.86 ± 0.07 &   -2.9 ± 0.07 \\
 $\alpha_{BB}$ &  -3.02 ± 0.23 &   -2.85 ± 0.20 &  -2.75 ± 0.22 &   -2.65 ± 0.20 &   -2.90 ± 0.31 &  -2.51 ± 0.23 \\ [1ex]
  \rule{0pt}{3ex}
$\beta_{EE}$  &  -3.22 ± 0.09 &  -3.14 ± 0.11 &   -3.13 ± 0.10 &    -3.00 ± 0.10 &  -3.01 ± 0.11 &   -2.98 ± 0.10 \\
 $\beta_{BB}$ &  -3.48 ± 0.41 &  -3.15 ± 0.41 &  -2.85 ± 0.38 &  -3.05 ± 0.36 &  -2.66 ± 0.55 &  -2.39 ± 0.39 \\ [1ex]
 \rule{0pt}{3ex}
 $A^{EE}$ $[10^{-3}\mu K]$ &   4.55 ± 0.18 &    6.79 ± 0.30 &   7.82 ± 0.33 &    9.18 ± 0.40 &  10.88 ± 0.47 &   12.84 ± 0.50 \\
 $A^{BB}$ $[10^{-3}\mu K]$ &    0.98 ± 0.20 &   1.47 ± 0.25 &   2.04 ± 0.31 &   2.21 ± 0.32 &    2.26 ± 0.50 &   3.67 ± 0.51 \\
 $A^{BB}/A^{EE}$ &   0.22 ± 0.04 &   0.22 ± 0.04 &   0.26 ± 0.04 &   0.24 ± 0.04 &   0.21 ± 0.05 &   0.29 ± 0.04 \\ [1ex]
 \rule{0pt}{2.5ex}
 $\chi_{EE}^2$ (41 dof) &        44.3 &        44.1 &         50.3 &        49.7 &        44.8 &        38.7 \\
$\chi_{BB}^2$ (41 dof) &        79.7 &        51.5 &        64.1 &        45.9 &         82.6 &        68.5 \\ [1ex] 
 \hline
\end{tabular}}
\caption{\textit{Planck}-\textit{WMAP} results. Best-fit parameters (with 1$\sigma$ errors) for the model given in equation \ref{eq:fit_beta} for the EE and BB power spectra, 
and their corresponding $\chi^2$ values. \textit{Planck} spectra are computed cross-correlating the half-ring maps, \textit{WMAP} spectra are computed cross-correlating the co-added \textit{WMAP} K-band years maps. The sky fractions are the ones retained by the six sky masks described in section \ref{sec:03}.}
\label{table:Cross_beta}
\end{table}

%% file: sections/section06.tex
\section{Conclusions}
\label{sec:06}

We have analyzed the sky emission at 23 and 30~GHz with the \textit{WMAP} K-band and the \textit{Planck} lowest frequency channel data. The main target of our analysis has been the study of the angular and spectral distribution of the diffuse Galactic polarized synchrotron. We have constructed a set of six masks, five of them increasing from low to intermediate  Galactic latitude (from 30 to 70 per cent of sky coverage), and a 94\% mask which allows almost the full sky except for the Galactic center and some bright point sources. We have estimated EE, BB and EB power spectra from \textit{Planck} and \textit{WMAP} independently, as well as by cross-correlating the two experimental data. We have fitted the power law $C_\ell^{EE,BB}\propto \ell^{\alpha_{EE,BB}}$ independently for the EE and BB power spectra and the constant $C_\ell^{EB} = A^{EB}$ for the EB cross-spectrum, in the multipole range 30 $\leq$ $\ell$ $\leq$ 300 for each of the considered cases.

For the cross-correlation analysis and a mask that allows 50 per cent of the sky, we have found a steep decay for E and B-modes, with indices $\alpha_{EE}= -2.95\pm0.04$ and $\alpha_{BB}=-2.85\pm0.14$, consistent between both components, and an asymmetry between the two amplitude modes with a B-to-E ratio equal to 0.22$\pm$0.02, at the pivot multipole $\ell = 80$. The compatibility between the two polarization components is better, in general, when considering mainly regions with high signal-to-noise. For the cross-correlation analysis, we have also found that the EB cross-spectra is consistent with zero at 1$\sigma$ for all the considered sky fractions, imposing a constraint on the EB amplitude to be $\le$ 1.2\% (2$\sigma$) that of the EE amplitude for the 50\% mask. We have also obtained, in general, consistent results from the \textit{Planck} and \textit{WMAP} independent analysis with respect to the ones found for the cross-correlation case. However, some small differences are present in the case of the best-fit parameters estimated only from the \textit{Planck} 30~GHz map. In particular, we find a less steep BB spectra ($\alpha_{BB}$ around 2.24) and a slightly larger $B/E$ ratio (around 0.27), even if consistency with the cross-correlation results holds at 2$\sigma$ in both cases.

We have done some robustness tests which have confirmed the validity of our results. In particular, we have fitted our model to the frequency maps of the 2020 \textit{Planck} \texttt{NPIPE} release (PR4), and in a larger multipole range (10 $\leq$ $\ell$ $\leq$ 400). Moreover, we have estimated spectra independently for the two hemispheres finding a larger emission in the North in the case of the E mode. Apart from that, there are not significant differences in the model of the polarization power spectra between the two hemispheres and with the full sky. 

We have fitted a simple power law to the synchrotron spectral energy distribution independently for the EE and BB spectra, considering both \textit{Planck} and \textit{WMAP} data. The recovered spectral indices $\beta_{EE}$ and $\beta_{BB}$, with respective values -3.00$\pm$0.10 and -3.05$\pm$0.36 for the 50 per cent mask, are compatible. The results indicate a trend of the spectral indices towards steeper values when higher Galactic latitudes are included in the analysis.

%% file: sections/appendix01.tex
\section{Mask Selection}
\label{appendix:01}

In this section, we present the procedure used to construct a reliable set of masks by imposing different threshold levels in polarization, such that the selected regions correlate well with those where the synchrotron emission has a higher signal-to-noise ratio. For this task, we consider two types of simulations: (i) only-foregrounds simulation at frequency 30 GHz, computed with the PySM model ("d1","s1"), (ii) simulation of \textit{Planck} data, as described in section \ref{sec:02}, adding the PySM foreground map, a CMB realization and a \textit{Planck} noise simulation.
From both simulated maps, once smoothed to $5^{\circ}$ resolution and after excluding the emission from the Galactic center and from point sources (as described in section \ref{sub:03,subsec:3}), we mask the total polarized intensity map below successively higher thresholds of $P$, selecting eight regions with $f_{sky}$ from 0.9 to 0.2 in steps of 0.1. We repeat the procedure with 5 different noise and CMB realizations. 

\begin{figure}[htbp]
\centerline{\includegraphics[scale=.5]{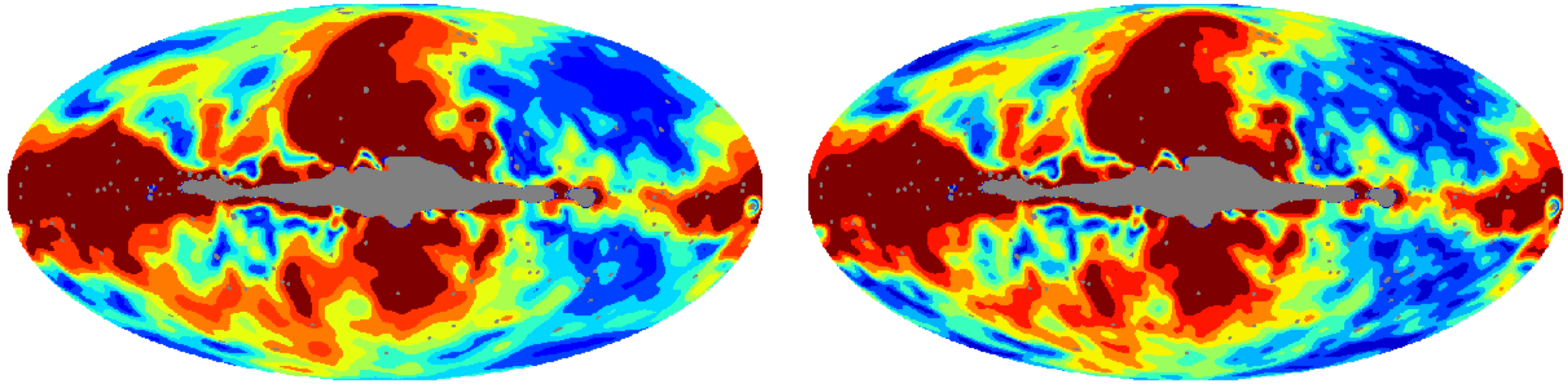}}
\caption{Left: All-sky map showing the sky regions computed with the successive thresholds applied to the PySM foregrounds simulation. Right: same regions for one simulation including the PySM foregrounds, CMB and noise.
The unmasked regions by the eight masks are showed from the smallest sky fraction mask (20\%) in dark-red, and by adding the regions in red, orange, yellow, green, turquoise, light-blue and blue, up to the one allowing the largest sky fraction (94\%) which leaves unmasked the full sky except for the grey region. This excluded region corresponds to the combination of the Galactic center mask and the point source mask.}
\label{fig:masks_sim}
\end{figure}

By comparing the masks constructed in this way from the only-foregrounds (that would provide the \emph{ideal} mask) and the complete simulations, we can see when the presence of other components is starting to affect the constructed mask and, therefore, at which threshold the selected regions do not correlate so well with the synchrotron amplitude.
For each realization we compute the cross-correlation coefficient $\rho$ between the only-foregrounds mask and the full-components mask. Moreover, we compute the foregrounds signal-to-noise ratio as the dispersion of the PySM foreground maps over the one of the CMB plus noise map, at scale of $1^\circ$, when the full-components mask is applied. The average values over the different realizations are reported in Table \ref{table:masks_sim}. \\
Comparing the regions allowed from the two mask sets shown in Figure \ref{fig:masks_sim}, we see that the masks constructed from the full simulated data start to be quite affected by noise for large sky fractions, deviating significantly from the masks constructed from the only-foregrounds simulation. The discrepancy is quantified by the cross-correlation and S/N values which, as expected, decrease with the sky fraction. 

From this insight, we decided not to consider in the analysis those masks with an average signal-to-noise ratio smaller than 2.5, to prevent the inclusion of too noisy regions. Moreover, we do not consider masks that retain a too small sky fraction, in order to limit the effect of the mask in the spectra estimation. Therefore, we pick the most reliable mask set as the one which retain a sky fraction ranging from 0.7 to 0.3. We select the mask with $f_{sky}$ = 50\% as the reference case for our main results, because it represents the best compromise of sky fraction, signal-to-noise ratio and cross-correlation between the ideal and realistic mask. 

The $f_{sky}$ = 0.94 mask that we use in the analysis, but not directly considered in this test, has the lowest signal-to-noise ratio and in some regions can be even dominated by noise, therefore, some considerations can be less reliable than for the other mask cases. However, we decided to show results also for the $f_{sky}$ = 0.94 mask in order to check if the characterization of the diffuse synchrotron features can be extended to the full sky, when the Galactic plane and bright point sources are properly masked.

\begin{table}[h]
\centering
\scalebox{0.8}{
\begin{tabular}{|c|c|c|}
\hline
\rule{0pt}{3ex}
$f_{sky}$ [\%] & $\rho$ &  $S/N$   \\ [0.5ex] 
\hline
\rule{0pt}{3ex}
\textbf{90}   & 0.82 & 2.37 \\
\textbf{80}   & 0.83 & 2.48 \\
\textbf{70}   & 0.88 & 2.60 \\
\textbf{60}   & 0.91 & 2.74 \\
\textbf{50}   & 0.92 & 2.88 \\
\textbf{40}   & 0.93 & 3.06 \\
\textbf{30}   & 0.93 & 3.27 \\
\textbf{20}   & 0.95 & 3.61 \\
\hline
\end{tabular}}
\caption{Cross-correlation and signal-to-noise values, corresponding to the different masks, estimated averaging over five only-foregrounds and all-components simulations (see text for details). The S/N value reported is the average between the Q and U signal-to-noise ratio. Note that differently to the masks used in the spectra estimation (see section \ref{sub:03,subsec:3}), these masks were not apodized and their boundaries were not regularized.}
\label{table:masks_sim}
\end{table}

%% file: sections/appendix02.tex
\section{Robustness test}
\label{appendix:02}
In order to test the robustness of our results we have carried out the following analyses.
First, following a similar procedure presented in section \ref{sec:04,subsec:2}, we test the power-law model cross-correlating the A/B detector splits of the new 2020 \textit{Planck} \texttt{NPIPE} release (PR4) maps at 30 GHz, in order to check the robustness of the results versus the considered 2018 \textit{Planck} release, which uses a different pipeline. Second, we fit the model to the same data set as in section \ref{sec:04,subsec:3}, but to a larger multipole range 10 $\leq$ $\ell$ $\leq$ 400, in order to check if the model holds when smaller and larger scales are included.

\subsection{Planck Release 4}
\label{appendix:02,subsection:01}
In 2020 the \textit{Planck} Collaboration has released new frequency maps in temperature and polarization using the \texttt{NPIPE} processing pipeline \cite{Npipe}. \texttt{NPIPE} introduces several improvements which lead to lower systematics as well as lower levels of noise, being the changes more significant for polarization data and for HFI channels.
Nevertheless, low frequency channels are also affected by the new pipeline and, therefore, it is worth checking the consistency of the results between the PR3 and PR4 releases. \\
The PR4 provides full-mission and A/B splits for data maps and simulations (for details see \cite{Npipe}). In particular, for the 30~GHz channel, the A subset is obtained combining maps of the years 1 and 3 and the B subset combining years 2, 4 and start of 5. In this section, we present the analysis performed cross-correlating the A/B splits of the \texttt{NPIPE} 30 GHz maps, degraded at the pixel resolution corresponding to $N_{side} = 512$.

We estimate the covariance matrices cross-correlating 300 A-split simulations with 300 B-split simulations provided by the \textit{Planck} Legacy Archive\footnote{pla.esac.esa.int}  (PLA).
Table~\ref{table:fitNpipecross} shows the best-fit parameters and the $\chi^2$ values for this case. These results are consistent within the errors with those found in section~\ref{sec:04,subsec:1} using \textit{Planck} release 3, as shown in Figure \ref{fig:Npipe-LR} (left panel) for the reference mask. As for PR3, spectra estimated from PR4 shows in general a less steep decay for both components, most notably for the B-mode, compared to the result found in \ref{sec:04,subsec:3} with the cross-analysis. However, the spectral indices $\alpha_{EE}$ and $\alpha_{BB}$ are, in general, more consistent between them for PR4 than for PR3.
The EB cross-term is also consistent with zero at 1$\sigma$ for the whole mask set. It is interesting to point out that the values found for the $\chi^2$ tend to be smaller when using PR4 data with respect to PR3, which is especially notable for the EB fit. This improvement is likely due to the larger number of simulations used in the PR4 analysis.
 \\

\begin{table}[h]
\centering
\scalebox{0.78}{
\begin{tabular}{|c|c|c|c|c|c|c|}
\hline
$f_{sky}$ & \textbf{94\%} &  \textbf{70\%} & \textbf{60\%} & \textbf{50\%} & \textbf{40\%} & \textbf{30\%} \\ [0.5ex] 
\hline
\rule{0pt}{3ex}
$\alpha_{EE}$ &  -2.86 ± 0.11 &  -2.82 ± 0.12 &  -2.81 ± 0.11 &  -2.84 ± 0.11 &  -2.87 ± 0.12 &  -2.96 ± 0.13 \\
$\alpha_{BB}$ &  -2.41 ± 0.23 &  -2.61 ± 0.19 &  -2.52 ± 0.17 &  -2.53 ± 0.24 &  -2.47 ± 0.28 &  -2.44 ± 0.35 \\ [1ex]
\rule{0pt}{3ex}
$A^{EE}$ $[10^{-3}\mu K]$ &   5.63 ± 0.31 &    8.0 ± 0.42 &   9.14 ± 0.47 &  10.34 ± 0.54 &  11.74 ± 0.62 &  14.37 ± 0.84 \\
$A^{BB}$ $[10^{-3}\mu K]$  &   1.74 ± 0.22 &    2.16 ± 0.2 &   2.36 ± 0.19 &   2.52 ± 0.27 &   2.94 ± 0.39 &   3.55 ± 0.54 \\
$A^{BB}/A^{EE}$ &   0.31 ± 0.04 &   0.27 ± 0.03 &   0.26 ± 0.02 &   0.24 ± 0.03 &   0.25 ± 0.04 &   0.25 ± 0.04 \\ [1ex]
\rule{0pt}{2.5ex}
$\chi_{EE}^2$ (20 dof) &        23.8 &        22.2 &        21.9 &        25.7 &        21.6 &        20.7 \\
$\chi_{BB}^2$ (20 dof) &        24.6 &        18.3 &         19.0 &        20.4 &         19.3 &        22.7 \\ [1ex] 
\hline
\rule{0pt}{3ex}
$A^{EB}$ $[10^{-3}º\mu K]$ &    0.04 ± 0.08 &     -0.02 ± 0.10 &    0.01 ± 0.11 &     -0.05 ± 0.1 &   -0.01 ± 0.12 &  0.0 ± 0.15 \\
$A^{EB}/A^{EE}$ &  0.007 ± 0.014 &  -0.002 ± 0.012 &  0.001 ± 0.012 &  -0.005 ± 0.009 &  -0.001 ± 0.010 &  0.0 ± 0.010 \\  [1ex]
\rule{0pt}{2.5ex}
$\chi_{EB}^2$ (21 dof) &  25.2 &   20.5 &  20.4 &   23.9 &  25.3 &  24.8 \\ [1ex]
\hline
\end{tabular}}
\caption{\textit{Planck} PR4 results. Best-fit parameters, 1$\sigma$ errors and $\chi^2$ values for the power-law in equation \ref{fit} for EE and BB, and for the constant baseline in \ref{fitEB} for EB. Power spectra are computed by cross-correlating A/B detector split of the \textit{Planck} \texttt{NPIPE} (PR4) 30~GHz maps, for each of the six sky masks described in section \ref{sec:03}.}
\label{table:fitNpipecross}
\end{table}

\subsection{Large Multipole Range}
\label{appendix:02,subsection:02}
In the main analysis we have considered the multipole range 30 $\leq$ $\ell$ $\leq$ 300. The upper limit is chosen because at higher multipoles both noise and possible emission of extra-Galactic compact sources can be important and then can strongly contaminate the foreground emission. The lower limit is chosen because 
pseudo-spectra methods (as \texttt{NaMaster}) are expected to be less reliable at small multipoles for masked sky regions. However, it makes sense to wonder if our results are robust when considering a larger range. Therefore, in this section we show the best parameters we find fitting equations \ref{fit}-\ref{fitEB} in the multipole range 10 $\leq$ $\ell$ $\leq$ 400 to the pseudo-$C_\ell$ computed cross-correlating the co-added 9 year \textit{WMAP} K-band maps and the full-mission \textit{Planck} (PR3) 30~GHz data, with exactly the same procedure described in section \ref{sec:04,subsec:3}. The fit parameters and $\chi^2$ values are reported in table \ref{table:fitCross_large}.

\begin{figure}[htp]
\centerline{\includegraphics[scale=.45]{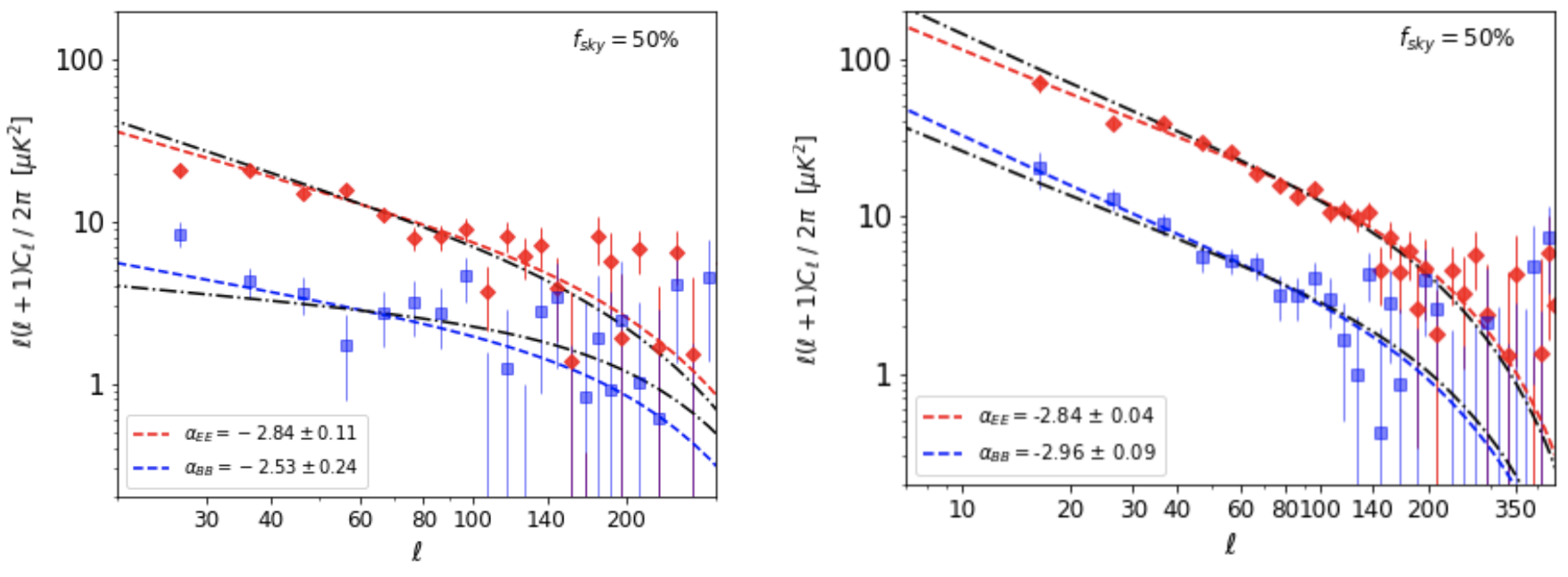}}
\caption{Best fit (dashed line) to the EE (red diamonds) and BB (blue squares) pseudo-spectra for the reference mask ($f_{sky} = 50\%$). Left: spectra are computed cross-correlating A/B splits of the \textit{Planck} \texttt{NPIPE} (PR4) 30~GHz maps. The fit with PR3 data presented in section \ref{sec:04,subsec:1} (black dash-dotted line) is given for comparison. Right: spectra are computed cross-correlating \textit{WMAP} K-band and the \textit{Planck} (PR3) 30~GHz maps and the fit is performed considering the large multipole range 10 $\leq$ $\ell$ $\leq$ 400. The best cross-analysis fit presented in section \ref{sec:04,subsec:3} (black dash-dotted line) is given for comparison.}
\label{fig:Npipe-LR}
\end{figure}

When working with the larger multipole range, we find in general slightly flatter values for both EE and BB for the different considered sky fractions, although this is not the case for the reference mask (see Fig.~\ref{fig:Npipe-LR}, right panel) where $\alpha_{BB}$ is actually slightly steeper.
Nevertheless, for the whole mask set the results are still quite compatible with those found in the main analysis. This indicates that the model is also valid at the larger scale range considered in this extended analysis.

\begin{table}[h]
\centering
\scalebox{0.78}{
\begin{tabular}{|c|c|c|c|c|c|c|}
\hline
$f_{sky}$ & \textbf{94\%} &  \textbf{70\%} & \textbf{60\%} & \textbf{50\%} & \textbf{40\%} & \textbf{30\%} \\ [0.5ex]
 \hline
   \rule{0pt}{3ex}
 $\alpha_{EE}$  &  -2.81 ± 0.04 &  -2.76 ± 0.04 &  -2.78 ± 0.05 &  -2.84 ± 0.04 &  -2.85 ± 0.05 &  -2.82 ± 0.06 \\
$\alpha_{BB}$   &  -3.12 ± 0.07 &  -3.05 ± 0.09 &  -2.96 ± 0.09 &  -2.96 ± 0.09 &  -2.85 ± 0.11 &  -2.87 ± 0.12 \\ [1ex]

  \rule{0pt}{3ex}
 $A^{EE}$ $[10^{-3}\mu K]$  &  10.39 ± 0.32 &  14.15 ± 0.38 &  16.05 ± 0.49 &   17.79 ± 0.5 &   20.78 ± 0.6 &  24.52 ± 0.87 \\
$A^{BB}$ $[10^{-3}\mu K]$  &   2.13 ± 0.13 &   2.97 ± 0.19 &   3.61 ± 0.22 &   3.94 ± 0.24 &   4.92 ± 0.33 &   5.95 ± 0.39 \\
$A^{BB}/A^{EE}$ & 0.20 ± 0.01 &   0.21 ± 0.01 &   0.23 ± 0.02 &   0.22 ± 0.02 &   0.24 ± 0.02 &   0.24 ± 0.02 \\ [1ex]
 \rule{0pt}{2.5ex}
 $\chi_{EE}^2$ (27 dof) &        44.9 &        31.7 &        38.1 &        34.3 &        31.2 &        48.1 \\
$\chi_{BB}^2$ (27 dof) &        19.1 &        18.7 &         20.0 &        17.4 &         23.6 &        21.2 \\ [1ex] 
 \hline
 \rule{0pt}{3ex}
 $A^{EB}$ $[10^{-3}\mu K]$ &    0.01 ± 0.07 &     0.12 ± 0.10 &     0.03 ± 0.10 &    0.02 ± 0.12 &    0.04 ± 0.11 &    0.01 ± 0.13 \\
 $A^{EB}/A^{EE}$ &  0.001 ± 0.007 &  0.008 ± 0.007 &  0.002 ± 0.006 &  0.001 ± 0.007 &  0.002 ± 0.005 &  0.001 ± 0.005 \\ [1ex]
 \rule{0pt}{2.5ex}
 $\chi_{EB}^2 $ (28 dof) &          60.1 &          70.6 &          52.5 &          53.9 &          37.1 &           30.4 \\ [1ex]
 \hline
\end{tabular}}
\caption{\textit{Planck}-\textit{WMAP} results. Best-fit parameters, 1$\sigma$ errors and $\chi^2$ values for the power-law in equation \ref{fit} for EE and BB, and for the constant baseline in \ref{fitEB} for EB. Power spectra is computed by cross-correlating the co-added 9 year \textit{WMAP} K-band maps and the full-mission \textit{Planck} 30~GHz maps, for each of the six sky masks described in section \ref{sec:03}.  Fits are performed on the multipole range 10 $\leq$ $\ell$ $\leq$ 400.}
\label{table:fitCross_large}
\end{table}

%% file: sections/appendix03.tex
\section{Hemisphere analysis}
\label{appendix:03}
\begin{figure}[htp]
\centerline{\includegraphics[scale=.58]{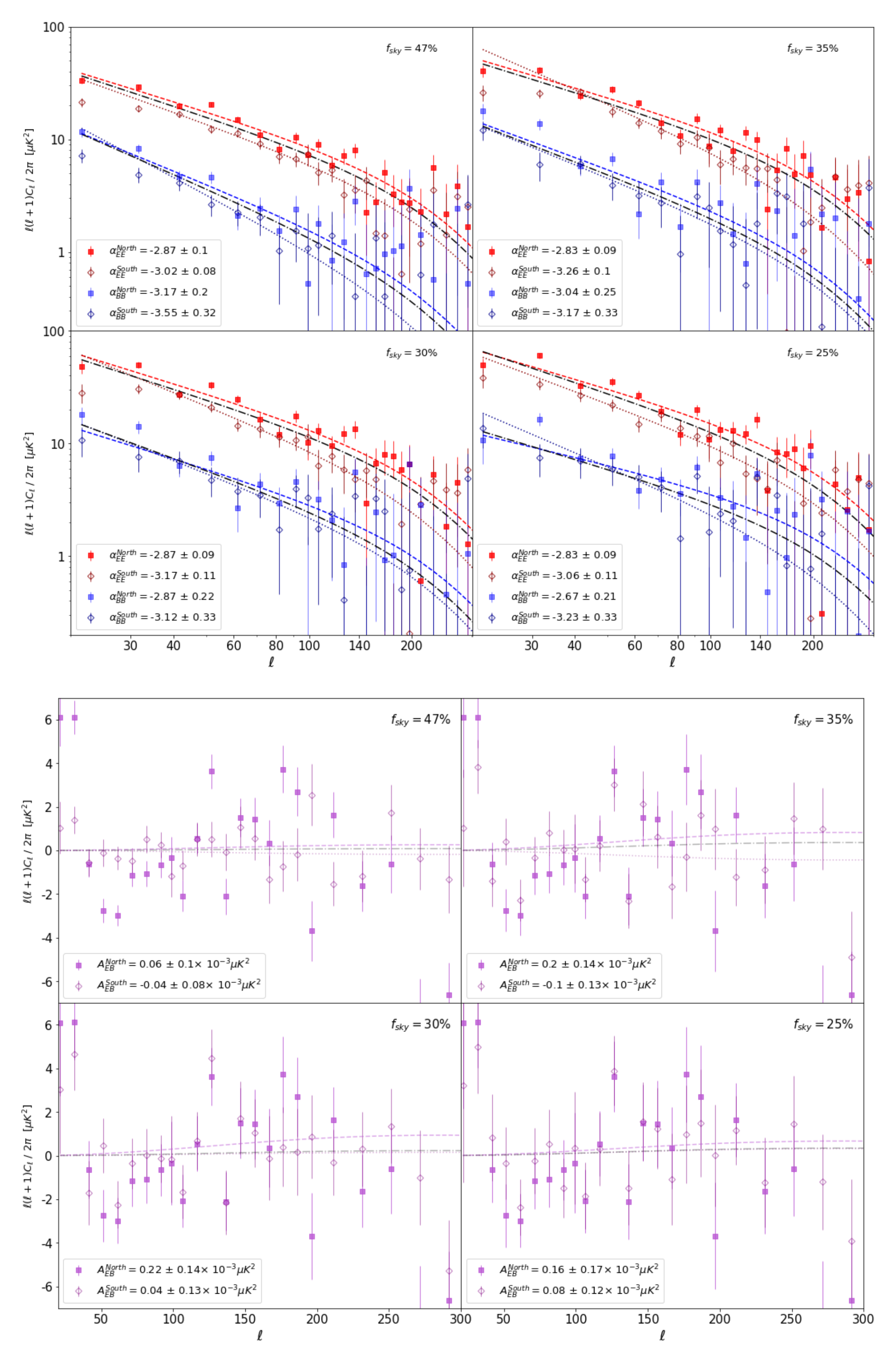}}
\caption{\textit{Planck}-\textit{WMAP} results. Top: Northern EE (red squares), Southern EE (
red diamonds), Northern BB (blue squares) and Southern BB (blue diamonds) pseudo-spectra. Bottom: Northern EB (purple squares) and Southern EB (purple diamonds) pseudo-spectra. Spectra are computed cross-correlating the co-added 9 year \textit{WMAP} K-band maps and the full-mission \textit{Planck} PR3 30~GHz maps, for the northern and southern parts of each of the four masks allowing the largest sky fractions. The $f_{sky}^S$ label of each panel indicates the area of the southern region allowed by the corresponding mask.
The dashed and dotted lines are, respectively, the Northern and Southern best fits to the hemisphere spectra and the black dash-dotted line is the best fit presented in section \ref{sec:04,subsec:3}.}
\label{fig:NS}
\end{figure}

In this appendix, we repeat the same analysis as in section~\ref{sec:04,subsec:3} but independently for the Northern and Southern hemispheres.
The mask set is the one presented in section \ref{sec:03}, where we simply separate regions from the two celestial hemispheres. We do not include the two most stringent masks in the analysis because they retain too small sky fractions which can negatively affect the spectra computation at low multipoles, where the diffuse synchrotron signal is important. We still keep the same multipole range (30 $\leq$ $\ell$ $\leq$ 300) and binning of the main analysis.
Table~\ref{table:NS} shows the best-fit parameters (with 1$\sigma$ errors) and $\chi^2$ values for both hemispheres while Figure~\ref{fig:NS} shows the power spectra and best-fit models for each case. For comparison, the best-fit model (black dot-dashed line) obtained from the full analysis is also shown, which tends to fall between the two hemisphere fits, showing a good level of consistency. 

\begin{figure}[tbp]
\centerline{\includegraphics[scale=.41]{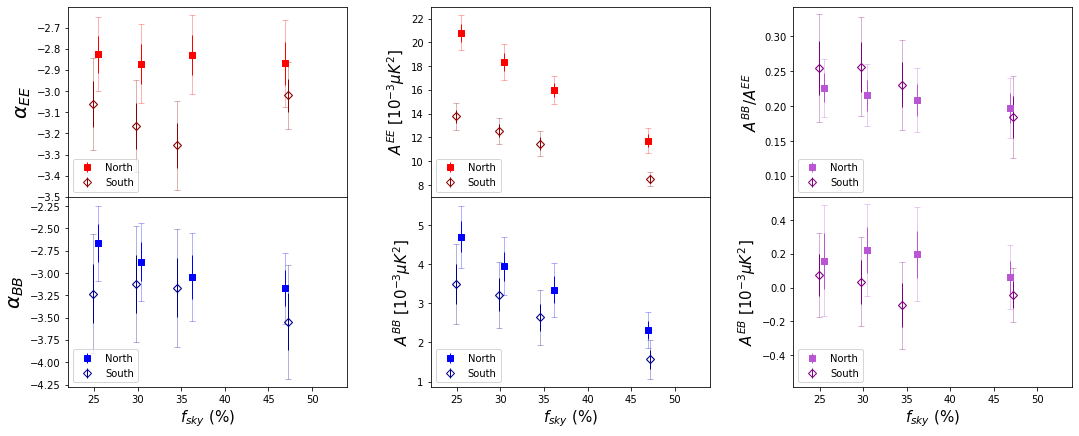}}
\caption{Comparison between the best-fit parameters found in the Northern hemisphere (squares) and in the Southern hemisphere (diamonds) to the models of equations \ref{fit}-\ref{fitEB}. 1$\sigma$ and 2$\sigma$ errors are showed, respectively, with thicker and thinner lines.}
\label{fig:North_vs_South}
\end{figure}

It is interesting to note that there are some differences between both hemispheres, as shown in Figure~\ref{fig:North_vs_South}, where the best-fit parameters for each hemisphere are compared. The synchrotron polarized emission in the Northern hemisphere is brighter than in the Southern hemisphere, with a factor around 1.4 larger for the amplitude of the EE spectra (slightly lower factor for BB). We also find a steeper decay of the synchrotron amplitude in the Southern hemisphere with respect to the Northern one. Nevertheless, the B-to-E ratio is quite consistent for the two hemispheres. The EB cross-term is compatible with zero at the 2$\sigma$ level for the whole mask set, even if the estimated EB/EE amplitude is  smaller for the Southern hemisphere. The goodness of the fits, in terms of the $\chi^2$ value, points out that the EE and BB power-law model with null EB term describes better the synchrotron polarization emission in the Southern hemisphere than in the Northern one. This discrepancy 
could be hinting that the mask procedure might be working better in the Southern than in the Northern hemisphere, where instead some complex structures, such as point sources or very bright Galactic plane emission, may remain unmasked. Nevertheless, the simple model considered in this analysis still seems to provide a reasonable good fit for both hemispheres. \\

\begin{table}[ht]
\centering
\scalebox{0.7}{
\begin{tabular}{|c|c|c|c|c|}
\hline
   \rule{0pt}{2.5ex}
$f_{sky}^N$ & \textbf{47\%} &  \textbf{36\%} & \textbf{30\%} & \textbf{26\%} \\ [0.5ex]
 \hline
   \rule{0pt}{3ex}
 $\alpha_{EE}$   &   -2.87 ± 0.10 &  -2.83 ± 0.09 &  -2.87 ± 0.09 &  -2.83 ± 0.09 \\
$\alpha_{BB}$   &   -3.17 ± 0.20 &  -3.04 ± 0.25 &  -2.87 ± 0.22 &  -2.67 ± 0.21 \\ [1ex]

  \rule{0pt}{3ex}
 $A^{EE}$ $[10^{-3}\mu K]$ &  11.74 ± 0.52 &  16.01 ± 0.58 &  18.36 ± 0.74 &   20.80 ± 0.74 \\
$A^{BB}$ $[10^{-3}\mu K]$  &   2.31 ± 0.23 &   3.34 ± 0.34 &   3.95 ± 0.37 &     4.70 ± 0.40 \\
$A^{BB}/A^{EE}$ &    0.20 ± 0.02 &   0.21 ± 0.02 &   0.22 ± 0.02 &   0.23 ± 0.02 \\ [1ex]
 \rule{0pt}{2.5ex}
 $\chi_{EE}^2$ (20 dof)&        45.0 &        33.9 &        37.9 &        30.9 \\
$\chi_{BB}^2$ (20 dof)  &        18.0 &        23.1 &        19.8 &        20.9 \\ [1ex] 
 \hline
 \rule{0pt}{3ex}
 $A^{EB}$ $[10^{-3}\mu K]$ &     0.06 ± 0.10 &     0.20 ± 0.14 &    0.22 ± 0.14 &    0.16 ± 0.17 \\
 $A^{EB}/A^{EE}$ &  0.007 ± 0.011 &  0.017 ± 0.012 &  0.018 ± 0.011 &  0.012 ± 0.012 \\ [1ex]
 \rule{0pt}{2.5ex}
 $\chi_{EB}^2 $ (21 dof)  &          31.4 &          40.4 &           30.5 &          35.5 \\ [1ex]
 \hline
\end{tabular}
}
\vspace{3mm}

\scalebox{0.7}{
\begin{tabular}{|c|c|c|c|c|}
\hline
   \rule{0pt}{2.5ex}
$f_{sky}^S$ & \textbf{47\%} &  \textbf{35\%} & \textbf{30\%} & \textbf{25\%}  \\ [0.5ex]
 \hline
   \rule{0pt}{3ex}
 $\alpha_{EE}$  &  -3.02 ± 0.08 &   -3.26 ± 0.1 &  -3.17 ± 0.11 &  -3.06 ± 0.11 \\
$\alpha_{BB}$  &  -3.55 ± 0.32 &  -3.17 ± 0.33 &  -3.12 ± 0.33 &  -3.23 ± 0.33 \\ [1ex]
  \rule{0pt}{3ex}
 $A^{EE}$ $[10^{-3}\mu K]$  &   8.51 ± 0.31 &  11.48 ± 0.52 &  12.55 ± 0.55 &  13.77 ± 0.57 \\
$A^{BB}$ $[10^{-3}\mu K]$  &   1.57 ± 0.24 &   2.64 ± 0.35 &   3.22 ± 0.43 &   3.51 ± 0.51 \\
$A^{BB}/A^{EE}$ &   0.18 ± 0.03 &   0.23 ± 0.03 &   0.26 ± 0.04 &   0.25 ± 0.04 \\ [1ex]
 \rule{0pt}{2.5ex}
 $\chi_{EE}^2$ (20 dof) &        14.7 &        19.1 &         17.8 &        16.3 \\
$\chi_{BB}^2$ (20 dof)  &        16.0 &        21.4 &        21.0 &        18.5 \\ [1ex] 
 \hline
 \rule{0pt}{3ex}
 $A^{EB}$ $[10^{-3}\mu K]$ &    -0.04 ± 0.08 &   -0.10 ± 0.13 &    0.04 ± 0.13 &    0.08 ± 0.12 \\
 $A^{EB}/A^{EE}$ &  -0.005 ± 0.009 &  -0.009 ± 0.011 &  0.003 ± 0.010 &  0.006 ± 0.009 \\ [1ex]
 \rule{0pt}{2.5ex}
 $\chi_{EB}^2 $ (21 dof) &            18.9 &          35.0 &           28.7 &          21.2 \\ [1ex]
 \hline
\end{tabular}
}
\caption{\textit{Planck}-\textit{WMAP} results. Top: Northern hemisphere, bottom: Southern hemisphere. Best-fit parameters with 1$\sigma$ errors and $\chi^2$ values of the power-law in equation \ref{fit} for EE and BB, and of the constant baseline in \ref{fitEB} for EB, computed cross-correlating the co-added 9 year \textit{WMAP} K-band maps and the full-mission \textit{Planck} 30~GHz maps. The masks used are constructed isolating Northern and Southern regions for the four masks (from 0.94 to 0.5) described in section \ref{sec:03}.}
\label{table:NS}
\end{table}

%% file: main.bbl
\providecommand{\href}[2]{#2}\begingroup\raggedright\begin{thebibliography}{10}

\bibitem{Bennett_2013}
C.L.~Bennett, D.~Larson, J.L.~Weiland, N.~Jarosik, G.~Hinshaw, N.~Odegard
  et~al., \emph{Nine-year wilkinson microwave anisotropy probe (wmap)
  observations: Final maps and results},
  \href{https://doi.org/10.1088/0067-0049/208/2/20}{\emph{The Astrophysical
  Journal Supplement Series} {\bfseries 208} (2013) 20}.

\bibitem{planckLegacy}
N.~Aghanim, Y.~Akrami, F.~Arroja, M.~Ashdown, J.~Aumont, C.~Baccigalupi et~al.,
  \emph{Planck 2018 results},
  \href{https://doi.org/10.1051/0004-6361/201833880}{\emph{Astronomy \&
  Astrophysics} {\bfseries 641} (2020) A1}.

\bibitem{HuWhite_1997}
W.~Hu and M.~White, \emph{Cmb anisotropies: Total angular momentum method},
  \href{https://doi.org/10.1103/physrevd.56.596}{\emph{Physical Review D}
  {\bfseries 56} (1997) 596–615}.

\bibitem{HuWhite_1996}
W.~Hu and M.~White, \emph{A new test of inflation},
  \href{https://doi.org/10.1103/physrevlett.77.1687}{\emph{Physical Review
  Letters} {\bfseries 77} (1996) 1687–1690}.

\bibitem{HuWhite_1996_2}
W.~Hu and M.~White, \emph{Acoustic signatures in the cosmic microwave
  background}, \href{https://doi.org/10.1086/177951}{\emph{The Astrophysical
  Journal} {\bfseries 471} (1996) 30–51}.

\bibitem{SpergelZald_1997}
D.N.~Spergel and M.~Zaldarriaga, \emph{Cosmic microwave background polarization
  as a direct test of inflation},
  \href{https://doi.org/10.1103/physrevlett.79.2180}{\emph{Physical Review
  Letters} {\bfseries 79} (1997) 2180–2183}.

\bibitem{CompSep2016}
R.~Adam, P.A.R.~Ade, N.~Aghanim, M.I.R.~Alves, M.~Arnaud, M.~Ashdown et~al.,
  \emph{Planck 2015 results},
  \href{https://doi.org/10.1051/0004-6361/201525967}{\emph{Astronomy \&
  Astrophysics} {\bfseries 594} (2016) A10}.

\bibitem{akrami2020planck}
Y.~Akrami, M.~Ashdown, J.~Aumont, C.~Baccigalupi, M.~Ballardini, A.J.~Banday
  et~al., \emph{Planck 2018 results-iv. diffuse component separation},
  \href{https://doi.org/10.1051/0004-6361/201833881}{\emph{Astronomy \&
  Astrophysics} {\bfseries 641} (2020) A4}.

\bibitem{Leach_2008}
S.M.~Leach, J.-F.~Cardoso, C.~Baccigalupi, R.B.~Barreiro, M.~Betoule, J.~Bobin
  et~al., \emph{Component separation methods for the planck mission},
  \href{https://doi.org/10.1051/0004-6361:200810116}{\emph{Astronomy \&
  Astrophysics} {\bfseries 491} (2008) 597–615}.

\bibitem{Dickinson_2018}
C.~Dickinson, Y.~Ali-Haïmoud, A.~Barr, E.~Battistelli, A.~Bell, L.~Bernstein
  et~al., \emph{The state-of-play of anomalous microwave emission (ame)
  research}, \href{https://doi.org/10.1016/j.newar.2018.02.001}{\emph{New
  Astronomy Reviews} {\bfseries 80} (2018) 1–28}.

\bibitem{G_nova_Santos_2016}
R.~Génova-Santos, J.A.~Rubiño-Martín, A.~Peláez-Santos, F.~Poidevin,
  R.~Rebolo, R.~Vignaga et~al., \emph{Quijote scientific results – ii.
  polarisation measurements of the microwave emission in the galactic molecular
  complexes w43 and w47 and supernova remnant w44},
  \href{https://doi.org/10.1093/mnras/stw2503}{\emph{Monthly Notices of the
  Royal Astronomical Society} {\bfseries 464} (2016) 4107–4132}.

\bibitem{planckDust2016}
R.~Adam, P.A.R.~Ade, N.~Aghanim, M.~Arnaud, J.~Aumont, C.~Baccigalupi et~al.,
  \emph{Planck intermediate results},
  \href{https://doi.org/10.1051/0004-6361/201425034}{\emph{Astronomy \&
  Astrophysics} {\bfseries 586} (2016) A133}.

\bibitem{planckDust}
{Planck Collaboration}, Y.~Akrami, M.~Ashdown, J.~Aumont, C.~Baccigalupi,
  M.~Ballardini et~al., \emph{Planck 2018 results: Xi. polarized dust
  foregrounds},
  \href{https://doi.org/10.1051/0004-6361/201832618}{\emph{Astronomy \&
  Astrophysics} {\bfseries 641} (2020) }.

\bibitem{Carretti_2019}
E.~Carretti, M.~Haverkorn, L.~Staveley-Smith, G.~Bernardi, B.M.~Gaensler,
  M.J.~Kesteven et~al., \emph{S-band polarization all-sky survey (s-pass):
  survey description and maps},
  \href{https://doi.org/10.1093/mnras/stz806}{\emph{Monthly Notices of the
  Royal Astronomical Society} {\bfseries 489} (2019) 2330–2354}.

\bibitem{C-BASS_2019}
M.A.~Stevenson, T.J.~Pearson, M.E.~Jones, C.J.~Copley, C.~Dickinson, J.J.~John
  et~al., \emph{The c-band all-sky survey (c-bass): digital backend for the
  northern survey}, \href{https://doi.org/10.1093/mnras/stz313}{\emph{Monthly
  Notices of the Royal Astronomical Society} {\bfseries 484} (2019)
  5377–5388}.

\bibitem{quijote12}
J.A.~{Rubi{\~n}o-Mart{\'\i}n}, R.~{Rebolo}, M.~{Aguiar},
  R.~{G{\'e}nova-Santos}, F.~{G{\'o}mez-Re{\~n}asco}, J.M.~{Herreros} et~al.,
  \emph{{The QUIJOTE-CMB experiment: studying the polarisation of the galactic
  and cosmological microwave emissions}},  in \emph{Ground-based and Airborne
  Telescopes IV}, vol.~8444 of \emph{Society of Photo-Optical Instrumentation
  Engineers (SPIE) Conference Series}, p.~84442Y, Sept., 2012,
  \href{https://doi.org/10.1117/12.926581}{DOI}.

\bibitem{FuskelandSynch}
U.~Fuskeland, K.J.~Andersen, R.~Aurlien, R.~Banerji, M.~Brilenkov, H.K.~Eriksen
  et~al., \emph{Constraints on the spectral index of polarized synchrotron
  emission from wmap and faraday-corrected s-pass data},
  \href{https://doi.org/10.1051/0004-6361/202037629}{\emph{Astronomy \&
  Astrophysics} {\bfseries 646} (2021) A69}.

\bibitem{Npipe}
Y.~Akrami, K.J.~Andersen, M.~Ashdown, C.~Baccigalupi, M.~Ballardini,
  A.J.~Banday et~al., \emph{Planck intermediate results},
  \href{https://doi.org/10.1051/0004-6361/202038073}{\emph{Astronomy \&
  Astrophysics} {\bfseries 643} (2020) A42}.

\bibitem{Planck2016Diff}
P.A.R.~Ade, N.~Aghanim, M.I.R.~Alves, M.~Arnaud, M.~Ashdown, J.~Aumont et~al.,
  \emph{Planck 2015 results},
  \href{https://doi.org/10.1051/0004-6361/201526803}{\emph{Astronomy \&
  Astrophysics} {\bfseries 594} (2016) A25}.

\bibitem{PySM}
B.~Thorne, J.~Dunkley, D.~Alonso and S.~Næss, \emph{The python sky model:
  software for simulating the galactic microwave sky},
  \href{https://doi.org/10.1093/mnras/stx949}{\emph{Monthly Notices of the
  Royal Astronomical Society} {\bfseries 469} (2017) 2821–2833}.

\bibitem{Haslam1981}
C.G.T.~Haslam, C.J.~Salter and H.~Stoffel, \emph{The all-sky 408 mhz survey},
  in \emph{Origin of Cosmic Rays}, G.~Setti, G.~Spada and A.W.~Wolfendale,
  eds., (Dordrecht), pp.~217--219, Springer Netherlands (1981),
  \href{https://doi.org/10.1007/978-94-009-8475-2_33}{DOI}.

\bibitem{Miville-Deschenes}
M.-A.~Miville-Deschenes, N.~Ysard, A.~Lavabre, N.~Ponthieu,
  J.~Mac{\'\i}as-Pérez, J.~Aumont et~al., \emph{Separation of anomalous and
  synchrotron emissions using wmap polarization data},
  \href{https://doi.org/10.1051/0004-6361:200809484}{\emph{åp} {\bfseries 490}
  (2008) 1093}.

\bibitem{Minami_2020}
Y.~Minami, \emph{Determination of miscalibrated polarization angles from
  observed cosmic microwave background and foreground eb power spectra:
  Application to partial-sky observation},
  \href{https://doi.org/10.1093/ptep/ptaa057}{\emph{Progress of Theoretical and
  Experimental Physics} {\bfseries 2020} (2020) }.

\bibitem{Planck_3}
N.~Aghanim, Y.~Akrami, M.~Ashdown, J.~Aumont, C.~Baccigalupi, M.~Ballardini
  et~al., \emph{Planck 2018 results},
  \href{https://doi.org/10.1051/0004-6361/201832909}{\emph{Astronomy \&
  Astrophysics} {\bfseries 641} (2020) A3}.

\bibitem{Planck2015_sim}
P.A.R.~Ade, N.~Aghanim, M.~Arnaud, M.~Ashdown, J.~Aumont, C.~Baccigalupi
  et~al., \emph{Planck 2015 results},
  \href{https://doi.org/10.1051/0004-6361/201527103}{\emph{Astronomy \&
  Astrophysics} {\bfseries 594} (2016) A12}.

\bibitem{healpy1}
A.~Zonca, L.~Singer, D.~Lenz, M.~Reinecke, C.~Rosset, E.~Hivon et~al.,
  \emph{healpy: equal area pixelization and spherical harmonics transforms for
  data on the sphere in python},
  \href{https://doi.org/10.21105/joss.01298}{\emph{Journal of Open Source
  Software} {\bfseries 4} (2019) 1298}.

\bibitem{healpy2}
K.M.~Gorski, E.~Hivon, A.J.~Banday, B.D.~Wandelt, F.K.~Hansen, M.~Reinecke
  et~al., \emph{Healpix: A framework for high‐resolution discretization and
  fast analysis of data distributed on the sphere},
  \href{https://doi.org/10.1086/427976}{\emph{The Astrophysical Journal}
  {\bfseries 622} (2005) 759–771}.

\bibitem{FilteredFusion}
F.~{Arg{\"u}eso}, J.L.~{Sanz}, D.~{Herranz}, M.~{L{\'o}pez-Caniego} and
  J.~{Gonz{\'a}lez-Nuevo}, \emph{{Detection/estimation of the modulus of a
  vector. Application to point-source detection in polarization data}},
  \href{https://doi.org/10.1111/j.1365-2966.2009.14549.x}{\emph{Monthly Notices
  of the Royal Astronomical Society} {\bfseries 395} (2009) 649}.

\bibitem{mhw2}
M.~{L{\'o}pez-Caniego}, D.~{Herranz}, J.~{Gonz{\'a}lez-Nuevo}, J.L.~{Sanz},
  R.B.~{Barreiro}, P.~{Vielva} et~al., \emph{{Comparison of filters for the
  detection of point sources in Planck simulations}},
  \href{https://doi.org/10.1111/j.1365-2966.2006.10639.x}{\emph{Monthly Notices
  of the Royal Astronomical Society} {\bfseries 370} (2006) 2047}.

\bibitem{WMAP_PS}
M.~López-Caniego, M.~Massardi, J.~González-Nuevo, L.~Lanz, D.~Herranz,
  G.~De~Zotti et~al., \emph{Polarization of the wmap point sources},
  \href{https://doi.org/10.1088/0004-637x/705/1/868}{\emph{The Astrophysical
  Journal} {\bfseries 705} (2009) 868–876}.

\bibitem{namaster}
D.~Alonso, J.~Sanchez and A.~Slosar, \emph{A unified pseudo-c$_\ell$
  framework}, \href{https://doi.org/10.1093/mnras/stz093}{\emph{Monthly Notices
  of the Royal Astronomical Society} {\bfseries 484} (2019) 4127}.

\bibitem{bilbaoahedo2021eclipse}
J.~Bilbao-Ahedo, R.~Barreiro, P.~Vielva, E.~Martínez-González and D.~Herranz,
  \emph{Eclipse: a fast quadratic maximum likelihood estimator for cmb
  intensity and polarization power spectra},
  \href{https://doi.org/10.1088/1475-7516/2021/07/034}{\emph{Journal of
  Cosmology and Astroparticle Physics} {\bfseries 2021} (2021) 034}.

\bibitem{Upham_2021}
R.E.~Upham, M.L.~Brown and L.~Whittaker, \emph{Sufficiency of a gaussian power
  spectrum likelihood for accurate cosmology from upcoming weak lensing
  surveys}, \href{https://doi.org/10.1093/mnras/stab522}{\emph{Monthly Notices
  of the Royal Astronomical Society} {\bfseries 503} (2021) 1999–2013}.

\bibitem{Azzoni_2021}
S.~Azzoni, M.~Abitbol, D.~Alonso, A.~Gough, N.~Katayama and T.~Matsumura,
  \emph{A minimal power-spectrum-based moment expansion for cmb b-mode
  searches}, \href{https://doi.org/10.1088/1475-7516/2021/05/047}{\emph{Journal
  of Cosmology and Astroparticle Physics} {\bfseries 2021} (2021) 047}.

\bibitem{SciPy}
P.~Virtanen, R.~Gommers, T.E.~Oliphant, M.~Haberland, T.~Reddy, D.~Cournapeau
  et~al., \emph{{{SciPy} 1.0: Fundamental Algorithms for Scientific Computing
  in Python}}, \href{https://doi.org/10.1038/s41592-019-0686-2}{\emph{Nature
  Methods} {\bfseries 17} (2020) 261}.

\bibitem{NicolettaSych}
N.~Krachmalnicoff, E.~Carretti, C.~Baccigalupi, G.~Bernardi, S.~Brown,
  B.M.~Gaensler et~al., \emph{S–pass view of polarized galactic synchrotron
  at 2.3 ghz as a contaminant to cmb observations},
  \href{https://doi.org/10.1051/0004-6361/201832768}{\emph{Astronomy \&
  Astrophysics} {\bfseries 618} (2018) A166}.

\bibitem{Planck_IS}
Y.~Akrami, M.~Ashdown, J.~Aumont, C.~Baccigalupi, M.~Ballardini, A.J.~Banday
  et~al., \emph{Planck 2018 results. vii. isotropy and statistics of the cmb},
  \href{https://doi.org/10.1051/0004-6361/201935201}{\emph{Astronomy \&
  Astrophysics} {\bfseries 641} (2020) A7}.

\bibitem{davies2006determination}
R.D.~Davies, C.~Dickinson, A.J.~Banday, T.R.~Jaffe, K.M.~Górski and
  R.J.~Davis, \emph{{A determination of the spectra of Galactic components
  observed by the Wilkinson Microwave Anisotropy Probe}},
  \href{https://doi.org/10.1111/j.1365-2966.2006.10572.x}{\emph{Monthly Notices
  of the Royal Astronomical Society} {\bfseries 370} (2006) 1125}.

\bibitem{ade2015joint}
P.~Ade, N.~Aghanim, Z.~Ahmed, R.~Aikin, K.~Alexander, M.~Arnaud et~al.,
  \emph{Joint analysis of bicep2/keck array and planck data},
  \href{https://doi.org/10.1103/physrevlett.114.101301}{\emph{Physical Review
  Letters} {\bfseries 114} (2015) }.

\bibitem{dickinson2016cmb}
C.~Dickinson, \emph{Cmb foregrounds - a brief review},  2016.

\bibitem{pltpy}
J.D.~Hunter, \emph{Matplotlib: A 2d graphics environment},
  \href{https://doi.org/10.1109/MCSE.2007.55}{\emph{Computing in Science
  Engineering} {\bfseries 9} (2007) 90}.

\bibitem{NumPy}
C.R.~Harris, K.J.~Millman, S.J.~van~der Walt, R.~Gommers, P.~Virtanen,
  D.~Cournapeau et~al., \emph{Array programming with {NumPy}},
  \href{https://doi.org/10.1038/s41586-020-2649-2}{\emph{Nature} {\bfseries
  585} (2020) 357–362}.

\end{thebibliography}\endgroup
